\documentclass[12pt,a4paper]{article}
\usepackage[english]{babel}

\usepackage[papersize={8.5in,11in}]{geometry}

\usepackage{mathrsfs} 
\usepackage{textcomp} 
\usepackage{bm} 

\usepackage{natbib}

\usepackage{indentfirst}

\pdfoutput=1

\usepackage{setspace}
\usepackage[in]{fullpage} 

\usepackage{times}
\usepackage{graphicx}
\usepackage{amsmath}

\usepackage{algorithm}
\usepackage[noend]{algpseudocode}

\usepackage{placeins}
\usepackage{setspace}
\usepackage{tabularx, booktabs}

\newcolumntype{Y}{>{\centering\arraybackslash}X}

\usepackage{float}

\newcommand{\vect}[1]{\boldsymbol{#1}}

\usepackage{amsmath}
\newcommand{\bigO}{\mathcal{O}}

\usepackage{gensymb}
\usepackage{makecell}
\usepackage{multirow}

\renewcommand{\cite}{\citep}

\DeclareMathOperator*{\argmin}{arg\,min}

\title{\vspace{-2cm} \textbf{Earthquake Arrival Association with \\ Backprojection and Graph Theory \vspace{-0.35cm}}} 

\author{Ian W. McBrearty$^1$, Joan Gomberg$^2$, Andrew A. Delorey$^1$, Paul A. Johnson$^1$}

\usepackage{amsmath}
\usepackage{amssymb}
\usepackage{breqn}
\usepackage{nameref}
\usepackage{gensymb}
\usepackage{makecell}

\usepackage{fancyhdr}

\date{}

\begin{document}

\maketitle
\vspace{-1.35cm}






\thispagestyle{fancy}
\fancyhf{} 
\renewcommand{\headrulewidth}{0pt}
\lfoot{\small{imcbrearty@lanl.gov}}
\rfoot{\small{LA-UR-19-21045}}
\cfoot{\thepage}

\begin{centering}

\vspace{0.5cm}

$^1$Geophysics Group, Los Alamos National Laboratory -- Los Alamos, NM, USA

$^2$United States Geological Survey -- Seattle, WA, USA

\end{centering}












\let\thefootnote\relax\footnote{Peer-Review disclaimer: This draft manuscript is distributed solely for purpose of scientific peer review. Its content is merely being considered for publication. Until the manuscript has been approved for publication by the U.S. Geological Survey (USGS), it does not represent any official USGS finding or policy.}



\vspace{-0.5cm}
\begin{abstract}

The association of seismic wave arrivals with causative earthquakes becomes progressively more challenging as arrival detection methods become more sensitive, and particularly when earthquake rates are high. For instance, seismic waves arriving across a monitoring network from several sources may overlap in time, false arrivals may be detected, and some arrivals may be of unknown phase (e.g., P- or S-waves). We propose an automated method to associate arrivals with earthquake sources and obtain source locations applicable to such situations. To do so we use a pattern detection metric based on the principle of backprojection to reveal candidate sources, followed by graph-theory-based clustering and an integer linear optimization routine to associate arrivals with the minimum number of sources necessary to explain the data. This method solves for all sources and phase assignments simultaneously, rather than in a sequential greedy procedure as is common in other association routines. We demonstrate our method on both synthetic and real data from the Integrated Plate Boundary Observatory Chile (IPOC) seismic network of northern Chile. For the synthetic tests we report results for cases with varying complexity, including rates of 500 earthquakes/day and 500 false arrivals/station/day, for which we measure true positive detection accuracy of $>$ 95$\%$. For the real data we develop a new catalog between January 1, 2010 - December 31, 2017 containing 817,548 earthquakes, with detection rates on average 279 earthquakes/day, and a magnitude-of-completion of $\sim$M1.8. A subset of detections are identified as sources related to quarry and industrial site activity, and we also detect thousands of foreshocks and aftershocks of the April 1, 2014 Mw 8.2 Iquique earthquake. During the highest rates of aftershock activity, $>$ 600 earthquakes/day are detected in the vicinity of the Iquique earthquake rupture zone.

\end{abstract}


\section*{INTRODUCTION}
\label{sec:Intro}

Earthquake catalogs are invaluable for monitoring seismic activity and for developing geophysical images of the Earth structure. Cataloging earthquakes generally follows the detection of impulsive arrivals, followed by the association of arrivals between stations to a common source, and then inferring the location, origin time, and magnitude of each earthquake. As improved instrumentation and increasing quantities of data have become available, arrival picking algorithms have markedly progressed \cite{gibbons2006detection, yoon2015earthquake, poiata2016multiband, perol2018convolutional, ross2018generalized, zhu2018phasenet}. Along with these improvements the number of detected arrivals generally increases approximately by an order of magnitude per decreasing magnitude unit of sensitivity, introducing new challenges into the earthquake detection pipeline. Converting high rates of arrival picks into an accurate earthquake catalog can be invaluable in seismology, since dense catalogs improve our understanding of seismogenic processes occurring at plate boundaries \cite{kato2014multiple, delorey2015cascading}, allow for monitoring rate changes of seismicity \cite{montoya2017bayesian, fiedler2018multiple}, improve resolution of tomographic images \cite{peng2005spatiotemporal, watkins2018local}, reveal dynamic triggering and anthropogenic induced seismicity \cite{shapiro2006hydraulic, hill2007dynamic,peng2009remote,ellsworth2013injection}, and may contain information regarding the timing of future  earthquakes \cite{rouet2017machine, lubbers2018earthquake}. Dense catalogs also provide new datasets that can be incorporated into increasingly popular machine learning approaches for a variety of applications in seismology \cite{devries2018deep, perol2018convolutional, ross2018generalized, zhu2018phasenet, zhu2018seismic}.

\begin{figure}[ht]
\centering
\includegraphics[width=0.98\linewidth]{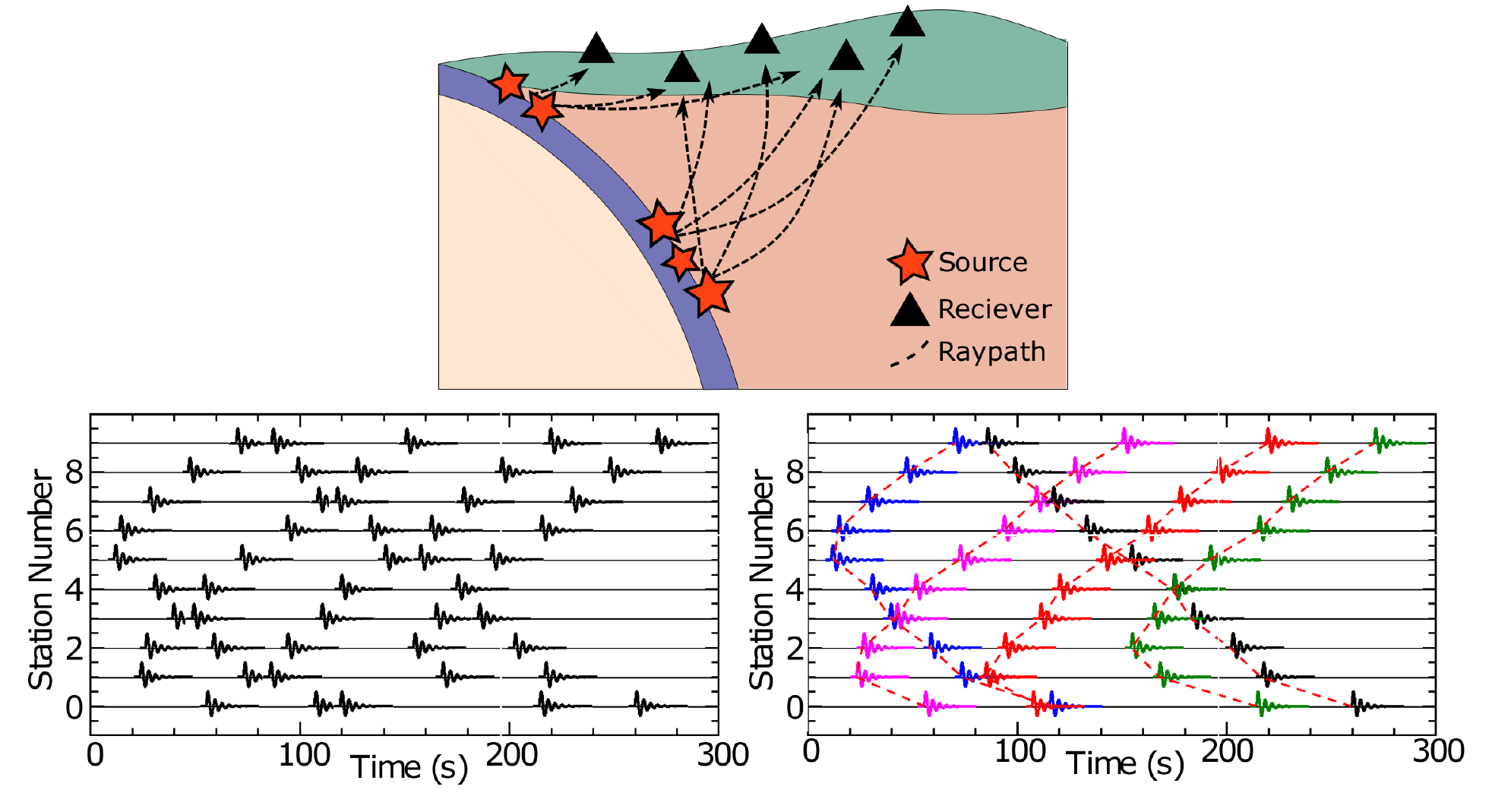}
\caption{Schematic example of the earthquake source-arrival association problem. (top) A subducting plate interface with several sources producing impulsive waves that propagate to nearby seismic stations. (bottom left) The observed set of unassociated arrivals (of a single phase type) across the seismic network. (bottom right) The correctly associated set of arrivals, colored into five distinct sources and linked by a moveout curve (red dashed line).}
\label{fig:Schematic}
\end{figure}

Existing earthquake association algorithms often struggle when arrival rates are high and there is ambiguity regarding the number of earthquakes that produce the arrival time measurements. False arrivals and unknown phase types of arrivals also introduce uncertainties. These issues have historically been most problematic during aftershock and foreshock sequences, however in the new era of sensitive arrival picking methods, these same issues are encountered for even normal background seismicity rates. As such, there is a need for improved association algorithms to enable the development of high fidelity low-magnitude-of-completion earthquake catalogs. A general purpose association algorithm should be able to handle a few inter-related problems, all connected to the key problem of earthquake detection. These include (1) an unknown number of sources in a time window, (2) false arrivals, (3) missing arrivals, (4) unknown phase types (e.g., P- or S-waves), (5) uncertainties in the velocity model, and (6) the scenario in which arrivals from several earthquakes overlap in time across a network (Fig. \ref{fig:Schematic}).

In this paper we solve the association problem when all of these uncertainties (1 - 6) are present using a backprojection and graph theory framework. The method processes a dataset of arrival time picks assumed to be of unknown phase, and which can include an unknown number of false arrivals and errors in arrival time estimates. The technique is generalizable to arbitrary spatial scales (from laboratory to Earth) and recording network geometries. In order to perform backprojection we rely on a velocity model, however the method can tolerate uncertainty in the velocity model and travel time calculations used in the analysis. In the process of inferring associations, estimates for hypocenter locations and origin times are also determined, which can be used as starting points for higher-precision methods. Effort has been taken to maintain a computationally efficient and scalable algorithm that can be used for catalog development in a wide range of applications. In conjunction with this work we also introduce a newly developed phase picker, referred to as the `Norm Picker', a single station energy-based picking algorithm that is similar to an STA/LTA \cite{allen1978automatic}, but that can pick P- and S-waves nearly equally as well and also pick low-SNR arrivals.

The structure of this paper is as follows: (1) Introduction and background, (2) Methods, (3) Synthetic Tests, (4) Application, (5) Discussion and (6) Conclusion. We perform synthetic tests of our method on simulated arrival time data from 17 stations of the Integrated Plate Boundary Observatory Chile (IPOC) of northern Chile (Table S1), and use real data from the same network for our application. In the synthetic tests, we analyze the performance of the method as the rate of earthquakes, the rate of false arrivals, and errors on arrival time data all change. We apply the method up to the relatively challenging synthetic case where 700 earthquakes/day, and 700 false arrivals/station/day occur across the network. For our application to real seismic data, we develop a catalog for northern Chile from 2010 through the end of 2017, which has a magnitude-of-completion of $\sim$M1.8 and detects on average 279 earthquakes/day.

\subsection*{Background}
\label{subsec:Background}

Previous studies have solved the association problem in several different ways with varying degrees of generalization and computational complexity. Several approaches use heuristic-based search procedures that involve choosing a set of arrivals occurring nearby in time, removing arrivals with high misfit to the inferred best-fitting location to these picks, and iteratively adding and removing new trial arrivals until convergence with a stable set of arrivals is obtained \cite{sippl2013geometry, gibbons2016iterative}. These techniques generally suffer from high computational cost and can scale with the squared or factorial of the number of arrivals considered \cite{johnson1997robust}. Using waveforms (rather than arrival times), network template matching can be used to resolve the association step implicitly \cite{gibbons2006detection}. Drawbacks of template matching include that templates must be known \textit{a priori}, there is a high computational cost, and generally only earthquakes that are similar to a template are detected. A related class of techniques transform raw seismic time-series into characteristic functions such as analytical envelopes or STA/LTA traces, before correlating the characteristic functions between stations to detect when a common source moveout crosses the network (for a review, see \citealt{li2017systematic}). Rule-based, fuzzy-logic, and `expert systems' have also been introduced \cite{bache1993knowledge}, as well as a Bayesian-based probabilistic association method \cite{arora2013net}. Other recent developments in earthquake association include a waveform-based pairwise association technique built around the Fingerprint and Similarity Thresholding (FAST) method \cite{bergen2018detecting}, a method using deep recurrent neural networks applied to arrival time data \cite{ross2019phaselink}, and a convolutional neural network approach for waveform-based pairwise associations \cite{mcbrearty2019pairwise}.

Backprojection, beamforming, and source-scanning are other methods for detecting earthquakes and implicitly associating arrivals \cite{kao2004source, de2015inverse, nakata2016reverse}. While details vary, in general these methods back-propagate waveforms from several stations into the source domain, and locations with high coalescence are identified as probable sources. Some variants replace full waveforms with finite-width kernels such as box-cars or scaled Gaussians centered about arrival times prior to implementing backprojection \cite{ringdal1989multi}. While effective in many scenarios, backprojection-based techniques generally yield difficult to interpret results when event rates are high and earthquakes occur closely together in time. In our work, we seek to resolve this ambiguity by interpreting backprojection results in a graph theory context. We use the remainder of this section to describe backprojection (and its subtle nuances) in greater detail.

\subsection*{Notation Conventions}

We define some notation conventions for clarity. The physical source space is given by $\Omega_{\vect{X}} \subseteq \mathbb{R}^3$, the origin time space is given by $\Omega_t \subseteq \mathbb{R}$, and the combined spatial-temporal domain by $\Omega = \Omega_{\vect{X}} \times \Omega_t$. The travel time to station $i$ from coordinate $\vect{X}$ of phase type $r$ is given by $T_i^r(\vect{X})$; the relative travel time with respect to the network is given by $\hat{T}_i^r(\vect{X}) = T_i^r(\vect{X}) - \min_{i,r} T_i^r(\vect{X})$. The set of all travel time vectors is given by $\mathcal{T} = \{T_i^r(\vect{X}) : \vect{X} \in \Omega_{\vect{X}}\}$; the set of all relative travel time vectors by $\hat{\mathcal{T}} = \{\hat{T}_i^r(\vect{X}) : \vect{X} \in \Omega_{\vect{X}}\}$. We let $\mathcal{D} = \{\tau_{i,j}\}$ represent the observed set of all $j$th arrivals on each $i$th station, and let $\tilde{\mathcal{D}}_{(\vect{X},t)} = \{t + T_i^r(\vect{X}) \}$ represent a set of theoretical arrivals recorded across the entire seismic network, for each $i$th station and $r$th phase type, from a source at $(\vect{X},t)$. A complete summary of the key notation used throughout the paper is given in Table 1.

\begin{table}[] \centering 

\caption{List of variables and their definitions.}

\begin{minipage}{0.95\linewidth}

\vspace{0.5cm}

\begin{tabularx}{\textwidth}{c c}
\cmidrule(lr){1-2}
 Symbol & \multicolumn{1}{l}{Description} \\
\cmidrule(lr){1-2}

$\vect{X}$ & \multicolumn{1}{l}{spatial coordinate (latitude, longitude, depth)} \\
$T_i(\vect{X})^r$ & \multicolumn{1}{l}{travel time to station $i$ from position $\vect{X}$ of phase type $r$} \\
$\hat{T}_i(\vect{X})^r$ & \multicolumn{1}{l}{relative travel time to station $i$ from position $\vect{X}$ of phase type $r$} \\
$C(\vect{X},t)$ & \multicolumn{1}{l}{backprojection space}  \\
$\tilde{C}_{(\tilde{\vect{X}})}(\vect{X},t)$ & \multicolumn{1}{l}{theoretical backprojection space for dataset of source $(\tilde{\vect{X}},t = 0)$}  \\
$\theta(\vect{X})$ & \multicolumn{1}{l}{prior likelihood distribution of sources}  \\
$\Omega$ & \multicolumn{1}{l}{spatial-temporal domain of sources} \\
$\Omega_{\vect{X}}$ & \multicolumn{1}{l}{spatial domain of sources} \\
$\Omega_{t}$ & \multicolumn{1}{l}{temporal domain of sources} \\
$\mathcal{T}, \mathcal{\hat{T}}$ & \multicolumn{1}{l}{set of all moveout vectors and set of all relative moveout vectors} \\
$\mathcal{R}$ & \multicolumn{1}{l}{residual distribution between template moveout vectors and moveout vectors} \\
$\mathcal{D}$ & \multicolumn{1}{l}{arrival time dataset} \\
$\tilde{\mathcal{D}}_{(\vect{X},t)}$ & \multicolumn{1}{l}{theoretical arrival time dataset of source $(\vect{X},t)$} \\
$\mathcal{S}$ & \multicolumn{1}{l}{candidate source dataset}  \\
$\mathcal{X}$ & \multicolumn{1}{l}{set of all template vector locations, $\vect{X}_k$ for $k = 1 \cdots P$} \\
$\tau_{i,j}$ & \multicolumn{1}{l}{$j$th arrival time of station $i$}  \\
$t_{k,l}$ & \multicolumn{1}{l}{$l$th triggering time of template $k$}  \\
$r_{obs}$ & \multicolumn{1}{l}{average rate of observed arrivals, per station, per day} \\
$w_{ijkl}^r$ & \multicolumn{1}{l}{edge-weights between $\tau_{i,j}$ arrival and $t_{k,l}$ triggering times (phase type $r$)} \\
$N, P$ & \multicolumn{1}{l}{$N$ stations, $P$ templates} \\
$A, b$ & \multicolumn{1}{l}{constraint matrix and constraint vector of competitive assignment} \\
$c, x$ & \multicolumn{1}{l}{optimization vector and solution vector of competitive assignment} \\
$\sigma, \gamma, \phi$ & \multicolumn{1}{l}{RBF-kernel width, triggering threshold, and cost term} \\
\cmidrule(lr){1-2}
\end{tabularx}

\end{minipage}
\end{table}

\subsection*{The Backprojection Space}
\label{subsec:backprojection}

Backprojection may be summarized as the reverse time-correction and stacking of recorded seismic waves by the relative theoretical travel time from a source to each station. When applied to a set of discrete arrival times from a seismic network, $\mathcal{D}$, let

\begin{equation} \label{eq:Backprojection}
    C(\vect{X},t) = \sum\limits_{\tau_{i,j} \in \mathcal{D}} \sum\limits_r f(t - (\tau_{i,j} - T^r_i(\vect{X})))
\end{equation}

\noindent denote the backprojection space (BP-space), $C(\vect{X},t)$, where $(\vect{X},t) \in \Omega$ represent possible spatial coordinates and origin times of sources. $f(\cdot)$ represents any type of locally supported kernel (centered about zero) used to map discrete arrivals onto continuous space and time; e.g., whenever $|t - (\tau_{i,j} - T^r_i(\vect{X}))| < \epsilon$ for small $\epsilon$, $f(\cdot)$ is non-zero and represents placing a kernel of some finite width along a manifold $\mathbb{M} \subset \Omega$, highlighting possible spatial-temporal sources of arrival $\tau_{i,j}$. The key feature of this mapping is that whenever a subset of arrivals in $\mathcal{D}$ are from true source-receiver(s) path(s) (e.g., $\tilde{\mathcal{D}}_{(\vect{X}_s,t_s)} \subseteq \mathcal{D} \; \text{for} \;  (\vect{X}_s,t_s) \in \Omega$), then $C(\vect{X},t)$ will have a significant degree of coalescence at $(\vect{X}_s,t_s)$, since several manifolds will all intersect and constructively interfere at this coordinate (i.e., in equation (\ref{eq:Backprojection}), all of the related terms $t_s - (\tau_{i,j} - T^r_i(\vect{X}_s)) = 0$ for $\tau_{i,j} \in \tilde{\mathcal{D}}_{(\vect{X}_s,t_s)}$). Note also that (\ref{eq:Backprojection}) is written for the general case where the directivity and phase types of arrivals are initially unknown, and each arrival is backprojected in all directions for all of the phase types under consideration.

\begin{figure}[]
\centering
\includegraphics[width=0.7\linewidth]{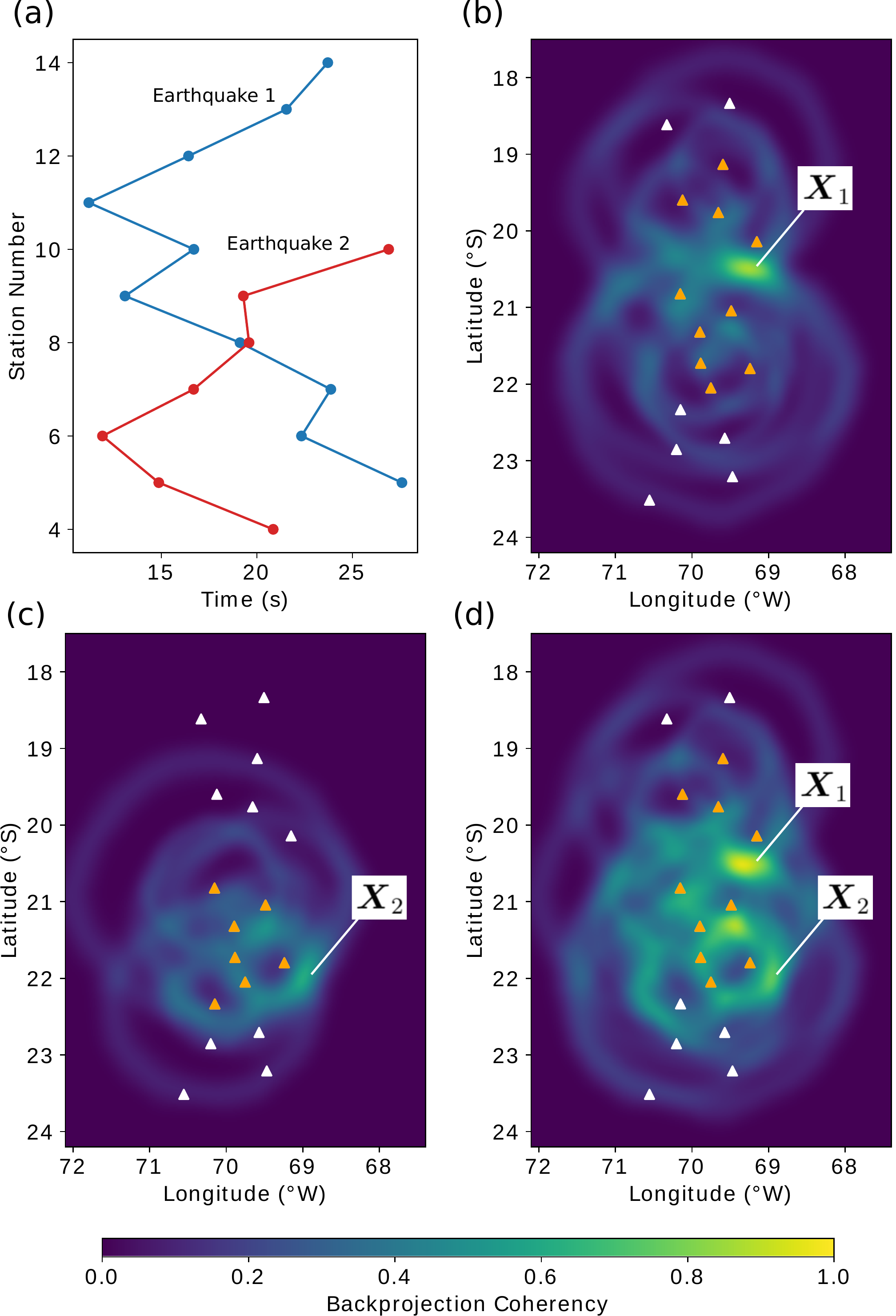}
\caption{Backprojection result for a two earthquake synthetic example. Both earthquakes occur at 50 km depth, 0 seconds origin time, and at positions marked $\vect{X}_1$ and $\vect{X}_2$ by white lines in panels (b,c,d). (a) The P-wave arrival times of two earthquakes recorded across the network, where $\vect{X}_1$ is recorded on ten stations and $\vect{X}_2$ is recorded on seven stations. (b,c) Backprojection results using only arrivals of $\vect{X}_1$, or $\vect{X}_2$, respectively. (d) Backprojection results using arrivals of both earthquakes. In (b,c,d) the BP-space is shown for 50 km depth and origin time of 0 seconds, and the RBF-kernel width used in equation (1) for backprojecting arrivals is $\sigma = $ 2 seconds.}
\label{fig:Backprojection}
\end{figure}

It is insightful to note that while the global maximum of $C(\vect{X},t)$ would identify the source location (and origin time) when $\mathcal{D} = \tilde{\mathcal{D}}_{(\vect{X}_s,t_s)}$, elsewhere throughout $\Omega$ a complex interference pattern is produced where subsets of the backprojected manifolds intersect. The coordinates of partial coalescence can form isolated, local maxima in $C(\vect{X},t)$, producing a topology in the BP-space that may be complex for even simple arrival time datasets of a single earthquake (Figs. \ref{fig:Backprojection}b,c). The complex interference patterns of the BP-space shown in Fig. \ref{fig:Backprojection} are typical for earthquakes in the interior of a seismic network. In teleseismic applications, or when the azimuthal coverage is poor, backprojected wavefronts are significantly more co-planar in the vicinity of the source location, reducing the number of subsidiary wavefront intersections.

As a revealing example, we consider a case where two earthquakes occur nearby in time, with one earthquake producing 10 P-wave arrivals, and the second earthquake producing 7 P-wave arrivals across the IPOC network (Fig. \ref{fig:Backprojection}a). Taken individually, equation (\ref{eq:Backprojection}) produces a BP-space with well defined global maxima (Figs. \ref{fig:Backprojection}b,c) at each of the respective source coordinates. When the arrival datasets are combined, because (\ref{eq:Backprojection}) is a linearly superposable field with respect to $\tau_{i,j}$, the interference patterns overlay one another and the identification of each individual source is more ambiguous (Fig. \ref{fig:Backprojection}d). The global maximum remains essentially invariant (from Fig. \ref{fig:Backprojection}b) at the location of the first source (with 10 arrivals), however the location of the second source - while remaining a local maximum - now has a coherency value on the order of many of the other local maxima in this space. In this case it's no longer evident whether or not it, or one of the other local maxima is a true source when using only the BP-space. This result is general: the coherency value of subsidiary modes (or local maxima) in the BP-space may not be diagnostic of which local maxima are true; e.g., a maxima in $C(\vect{X},t)$ may be a false source, yet have higher coherency than another maxima that is itself a true source. This later case is common when one earthquake has more arrivals than another nearby earthquake, since the secondary modes of the larger source may have higher coherency than even the primary maxima of the smaller source.  We also note that the BP image space maps shown in Fig. \ref{fig:Backprojection} are for the fixed, correct values of depth and origin time, and in practice the BP-space has four degrees of freedom (latitude, longitude, depth, and origin time), each of which introduces additional structure, local maxima, and variability.

\subsection*{The Backprojection Space for Theoretical Data}

One helpful insight into the ambiguity of the BP-space when multiple earthquakes are present in a short time-interval is that (\ref{eq:Backprojection}) predicts \textit{a priori} what the interference pattern will look like throughout all of $\Omega$ for any given choice of $\mathcal{D}$. This type of information has utility in determining which of many local maxima in $C(\vect{X},t)$ may be true, because it can indicate which set of local maxima are expected for any given source. If, for example, a source at $(\vect{X}_s,t_s)$ and its associated dataset $\tilde{\mathcal{D}}_{(\vect{X}_s,t_s)}$ has a predictably large coherency value at a local maxima $(\vect{X}_1,t_1) \neq (\vect{X}_s,t_s)$ underneath the mapping (\ref{eq:Backprojection}), we might be skeptical that the locally coherent point $(\vect{X}_1,t_1)$ is a true source when compared against the point of higher coherency, $(\vect{X}_s,t_s)$. Conversely, if a local maxima existed at $(\vect{X}_2,t_2) \neq (\vect{X}_s,t_s)$, where the BP-space is predictably low given $\tilde{\mathcal{D}}_{(\vect{X}_s,t_s)}$, we would have stronger evidence that $(\vect{X}_2,t_2)$ may be a second, true source, in addition to that of $(\vect{X}_s,t_s)$. 

This type of duality between the joint likelihood of pairs of sources will be used in a later step of our method. Analogously to the theoretical dataset, $\tilde{\mathcal{D}}_{(\vect{X}_s,t_s)}$, we define the theoretical backprojection space as

\begin{equation} \label{eq:TheoreticalCoherency}
    \tilde{C}_{(\tilde{\vect{X}})}(\vect{X},t) = \sum\limits_{\tau_{ij} \in \mathcal{\tilde{D}}}\sum\limits_r f(t - (\tau_{i,j} - T^r_i(\vect{X}))) \; ,
\end{equation}

\noindent which is equal to (\ref{eq:Backprojection}) computed for the case where $\mathcal{D} = \tilde{\mathcal{D}}_{(\tilde{\vect{X}}, t = 0)}$. We note that (\ref{eq:TheoreticalCoherency}) assumes all of the arrivals of theoretical source $(\tilde{\vect{X}}, t = 0)$ are contained in $\tilde{\mathcal{D}}_{(\tilde{\vect{X}},t = 0)}$, and also that (\ref{eq:TheoreticalCoherency}) is symmetric with respect to arbitrary pairs of sources $\{(\vect{X}_1,t_1)$, $(\vect{X}_2,t_2)\}$, in the sense that $\tilde{C}_{(\vect{X}_1)}(\vect{X}_2,t_2 - t_1) = \tilde{C}_{(\vect{X}_2)}(\vect{X}_1,t_1 - t_2)$ whenever the kernel $f(\cdot)$ is symmetric.

\section*{ASSOCIATION METHODS}
\label{sec:Methods}

The primary objective of our method is to process the BP-space produced by any of the complications considered (1 - 6 in the Introduction) under the mapping (\ref{eq:Backprojection}) and still infer the correct number of sources, the associations of arrivals to their correct source, and the identification of phase types and of false arrivals. The basic outline of our method is to detect coalescent points in the BP-space with an efficient search procedure. With these detections, we assemble a mathematical graph which connects all coalescent points (candidate sources) to arrivals consistent with those sources. We then formulate and apply a constrained linear optimization on the graph itself, which imposes physical constraints on the allowed connections (called `edges' in graph theory parlance) between candidate sources and arrivals. This step enables us to extract the true sources from the full graph and discard the false coalescent sources. In this process, arrivals are assigned to their optimal sources, phase types of arrivals are determined, and most false arrivals are discarded. For numerical efficiency, we also describe a graph pruning technique that can be applied prior to the optimization routine, which is useful to keep the optimization problem tractable and allow the method to scale over continuous time (streams of arrival times) applications. The pruning routine is implemented by combining information contained in both the observed and theoretical BP-space between pairs of sources into a mathematical graph that connects all sources to one another, and which allows for the rapid removal of the least optimal sources by using the Markov Clustering Algorithm \cite{dongen2000cluster}. All of the steps of the method are outlined in the flowchart titled Algorithm 1.

\subsection*{Template Moveout Vectors}
\label{subsec:Vector Quantization}

To detect all local maxima in $C(\vect{X},t)$ for an arbitrary dataset $\mathcal{D}$, one could theoretically compute $C$ over a dense grid-search of discretized $(\vect{X},t) \in \Omega$ and robustly find all local maxima. This process, however, would be computationally inefficient. A more practical alternative is to select a small, finite set of representative spatial backprojection coordinates, $\mathcal{X} = \{\vect{X}_k\}$, and compute $C(\vect{X}_k,t)$ for all $\vect{X}_k \in \mathcal{X}$, and over all time $t \in \Omega_t$. Since absolute seismic travel times to a fixed station change only gradually as source coordinates change, the space of all moveout vectors generally has low order complexity (i.e., varies smoothly), which ensures that a set of optimally chosen representative coordinates, $\{\vect{X}_k\}$, with their corresponding relative template travel time vectors, $\{\hat{T_i^r}(\vect{X}_k)\}$, can capture the broad characteristics of the space of all relative travel time vectors, $\hat{\mathcal{T}}$. Fig. S1 illustrates the essence of this partitioning approach, in which we have subdivided the region surrounding the IPOC seismic network into eight subregions, where for each subregion a single moveout vector approximates the true moveout vector for any source within that subregion. In this case eight subregions are shown for clarity, though in practice the optimal number is generally in the hundreds, which produce many more higher resolution sub-divisions than shown in Fig. S1.

To find a representative set of template vectors that optimally captures the main characteristics of $\hat{\mathcal{T}}$, we implement vector quantization \cite{vasuki2006review} with the K-Means clustering algorithm \cite{jain2010data}. K-Means is an efficient algorithm that can identify $P$ optimal cluster centroids (e.g., template vectors) by iteratively sampling vectors from $\hat{\mathcal{T}}$ and updating a set of template vectors, $v_k$ ($k = 1 \cdots P)$. The updates ensure that, statistically, arbitrary vectors in $\hat{\mathcal{T}}$ have at least one relatively similar template vector in $\{v_k\}$. That is, the algorithm updates the templates to iteratively minimize the distribution of nearest-neighbor residuals, 

\begin{equation}\label{eq:TemplateResidual}
    \mathcal{R} = \{ \; \min\limits_{k} \; || \; \hat{T}_i^r(\vect{X}) - v_k \; ||_{RMS} \; : \; \vect{X} \in \Omega \; \} \;
\end{equation}

\noindent between vectors in $\{v_k\}$ and those in $\hat{\mathcal{T}}$. As can be confirmed with empirical tests and by evaluating $\mathcal{R}$ for arbitrary choices of $\{v_k\}$, this type of clustering shows that $\{v_k\}$ can be chosen more optimally than simply uniformly sub-dividing space. This is because more cluster centroids can locate in areas with high travel time variability, while a sparse set of cluster centroids can occupy regions with highly similar sets of moveout vectors (for instance in areas with poor seismic network azimuthal coverage). In practice, the template moveout vectors returned from K-Means may not have an exact corresponding travel time vector from any given $\vect{X} \in \Omega_{\vect{X}}$. However, $\vect{X}_k$ can be taken as the nearest matching source coordinate to the template vectors; i.e., $\vect{X}_k = \argmin\limits_{\vect{X}} ||\hat{T}_i^r(\vect{X}) - v_k||_{RMS}$ for each $k = 1 \cdots P$.

In this application the minimum travel time was subtracted from each travel time vector prior to running K-Means, since $\hat{\mathcal{T}}$ has a simpler structure than $\mathcal{T}$, and only relative arrival times are important when applying backprojection techniques. We note that in practice $\hat{\mathcal{T}}$ can either be known on a discrete grid (such as the case for most 3D velocity model ray-tracing algorithms), or can be sampled from continuous space during the clustering algorithm's iterations if analytical formula of $T_i^r(\vect{X})$ are available. The fidelity with which the obtained template moveout vectors characterize all of space can be evaluated by estimating the residual distribution (\ref{eq:TemplateResidual}), which can help guide the choice of $P$ that should be used for any given application.

\subsection*{Discrete Backprojection with Template Moveout Vectors}
\label{subsec:Pattern Detection}

Rather than backproject to all of continuous $(\vect{X},t)$, for efficiency we backproject to the discrete templates located at $\vect{X}_k \in \mathcal{X}$, with the fixed template moveout vectors obtained from K-Means clustering. This procedure, we refer to as `discrete backprojection', can be succinctly written as a slight modification of equation (\ref{eq:Backprojection}):

\begin{subequations} \label{eq:Discrete Backprojection}
\begin{align}
C(\vect{X}_k,t) &= \frac{1}{N} \sum_{i = 1}^{N} \sum_{r = 1}^{2} \Bigl[ \sum_{j} f(t - (\tau_{i,j} - \hat{T}_i^r(\vect{X}_k))) \Bigr]^{\perp} \\
f(t) &= \frac{1}{2} exp \Bigl(\frac{-t^2}{2(\sigma^2 + \hat{\sigma}^2(\cdot))} \Bigr) \; \\
\text{for}& \; \tau_{i,j} \in \mathcal{D} \; \text{and} \; \vect{X}_k \in \mathcal{X} \; .
\end{align}
\end{subequations}

\noindent The discrete backprojection algorithm stacks the backprojected finitely supported radial basis function (RBF) kernel, $f(\cdot)$, with standard deviation $\sigma$, for all arrivals $\tau_{i,j} \in \mathcal{D}$, templates $\vect{X}_k \in \mathcal{X}$, and phases $r$. The number of stations is given by $N$, and the additional operator $[\cdot]^\perp$ is used to clip the maximum contribution to $C_k(t)$ of any given station to a maximum of $0.5$ for any given fixed template and phase type. This is useful to prevent an individual station from saturating $C_k(t)$ when, for example, the station has many closely spaced false arrivals in $\mathcal{D}$. Here, we have written (\ref{eq:Discrete Backprojection}) explicitly for the case of two-phases (e.g., P- and S-waves), and the scaling is designed such that whenever $\mathcal{D}$ contains a set of P- and S-wave arrivals recorded across all of the network equal to a template moveout vector, (i.e., $\{t_s + \hat{T}_i^r(\vect{X}_k) \} \subseteq \mathcal{D}$), then $C_k(t_s) = 1$. For completeness we have also included the optional term, $\hat{\sigma}(\cdot)$, in (\ref{eq:Discrete Backprojection}), to indicate that RBF-kernel widths can adapt for different pairs of source coordinates and stations if uncertainties in travel times vary across the seismic network and monitoring region.

It is straightforward to compute ($\ref{eq:Discrete Backprojection}$) for any arbitrary dataset $\mathcal{D}$. At this step in the method, we are interested in detecting all local maxima in $C(\vect{X}_k,t)$ (referred to as candidate sources), regardless if many of the maxima are not true sources. To achieve this we choose a hyperparameter, $\gamma$, referred to as the `triggering threshold', and detect all local peaks in $C_k(t) \geq \gamma$ using basic 1D peak-finding algorithms. While the maximum value (\ref{eq:Discrete Backprojection}) can obtain is 1, in general it's more practical to choose a relatively low $\gamma$. This will allow for the detection of sources at intermediate coordinates to any of the $\vect{X}_k$, detection of sources with noisy arrival time estimates (or uncertainties in travel time calculations), and also the detection of small magnitude sources that do not produce arrivals on all stations. Since each template has a characteristic source location, $\vect{X}_k \in \mathcal{X}$, the set of all local maxima in (\ref{eq:Discrete Backprojection}) can be assigned equivalent sources in $\Omega$; the set of all candidate sources is given by $\mathcal{S} = \{(\vect{X}_k,t_{k,l} - \min_{i,r} T_i^r(\vect{X}_k)) \}$, for each $l$th triggering time of the $k$th template, $t_{k,l}$. The time-offset $\min_{i,r} T_i^r(\vect{X}_k)$ from the triggering time $t_{k,l}$ is required since the template moveout vectors encode relative arrival times (i.e., have the minimum travel time subtracted), and the correction is needed to return the template triggering time back into the absolute time domain of origin times in $\Omega_t$.

Calculating (\ref{eq:Discrete Backprojection}) with a fixed kernel, $f(\cdot)$, and a finite set of template vectors (where usually it is sufficient to set $P < 1000$), makes discrete backprojection and the identification of candidate sources highly computationally efficient. This replaces the need for a dense grid-search over the entire source domain. However, a valid criticism may be that computing (\ref{eq:Discrete Backprojection}) only yields a sparse and sub-optimal sampling of $C(\vect{X},t)$, since in reality none of the sources in $\mathcal{S}$ are actually at maxima in the continuous $C(\vect{X},t)$ space (e.g., a maxima in $C(\vect{X}_k,t)$ restricted to $\vect{X} = \vect{X}_k$ may be on the flank of a true peak in continuous $\Omega$ space). Our experiments have shown that this effect on the remainder of our algorithm is relatively small as long as $P$ is large enough; however, to circumvent this issue we introduce an optional step that significantly reduces the resulting sensitivity to moveout templates. Namely, we take all candidate sources, $(\vect{X}_k,t_k - \min_{i,r} T_i^r(\vect{X}_k)) \in \mathcal{S}$, and update their locations in continuous $C(\vect{X},t)$ until a local maxima is reached by using standard local gradient ascent.  Such an approach is allowed since (\ref{eq:Discrete Backprojection}) is differentiable whenever $\partial T_i^r(\vect{X})/\partial\vect{X}$ is available. To account for the general case where $T_i^r(\vect{X})$ is known only on a discrete grid and $\partial T_i^r(\vect{X})/\partial\vect{X}$ is unknown, we propose fitting $T_i^r(\vect{X})$ with a simple fully-connected neural network that will ensure $T_i^r(\vect{X})$ is continuous and differentiable inside $\Omega$ (for a demonstration, see Supplemental Materials). In this framework, locally optimizing sources in $C(\vect{X},t)$ is efficient, especially when using automatic differentiation software (e.g., \citealt{paszke2017automatic}), and the computations can also be sped up with GPU acceleration.

\subsection*{Graph Construction of Source-Arrival Pairs}
\label{subsec:Graph Extraction}

Following discrete backprojection (and the optional local gradient ascent), a set of candidate sources, $\mathcal{S}$, has been obtained. As discussed in the Introduction, the relative coherency values of maxima in the BP-space are not themselves sufficient to determine which sources are real or false (Fig. \ref{fig:Backprojection}). Further, the BP-space has a strong interference pattern and a fixed (low) triggering threshold, $\gamma$, is used to identify all candidate sources. This results in only a small subset of the total number of candidate sources in $\mathcal{S}$ being true sources. The key component of our method is to make use of more information that is available than simply the coherency value of maxima in the BP-space. The main insight is that following the detection of all candidate sources, we know which (and by how much) each arrival contributed to the coherency value of each source. Namely, by inspection of (\ref{eq:Discrete Backprojection}), we see that for any pair $(\tau_{i,j},t_{k,l})$, the arrival $\tau_{i,j}$ contributed

\begin{equation}\label{eq:edge weights}
    w_{ijkl}^r = f(t_{k,l} - (\tau_{i,j} - \hat{T}_i^r(\vect{X}_k))) = \frac{1}{2} exp \Bigl(\frac{-(t_{k,l} - (\tau_{i,j} - \hat{T}_i^r(\vect{X}_k)))^2}{2(\sigma^2 + \hat{\sigma}^2(\cdot))} \Bigr)
\end{equation}

\noindent to $C(\vect{X}_k,t_{k,l})$ for phase type $r$\footnote{Technically, if the clipping operator is used in equation (\ref{eq:Discrete Backprojection}), then this is the value that would have been contributed prior to the clipping operator being used.}. Since this quantity can be computed for all pairs of arrival(s)-sources(s), the set of all $w_{ijkl}^r$ can also be used to represent the edge weights of a mathematical graph that links all source-arrival pairs by a weight proportional to how much each arrival contributed to the detection of each source (which is also dependent on the phase type $r$). A large edge weight quantity for a source-arrival (and phase dependent) pair indicates the arrival constructively interfered with several other backprojected arrivals to elucidate the source in question. Conversely, when the edge weight goes to zero, the arrival didn't contribute to the detection of the source with that phase type. Over continuous-time applications, a graph of this kind will be highly sparse, since each arrival will only be connected to a small subset of the total number of candidate sources. We also note that (\ref{eq:edge weights}) is written for the case where sources have not been locally optimized in $C(\vect{X},t)$ space, however if they have been, the only change is that $t_{k,l}$ becomes the new origin time estimate, and $\vect{X}_k$ becomes the new source coordinate in continuous $\Omega_{\vect{X}}$ (rather than being fixed at the original $t_{k,l}$ and $\vect{X}_k$ values).

\begin{figure}[]
\centering
\includegraphics[width=0.99\linewidth]{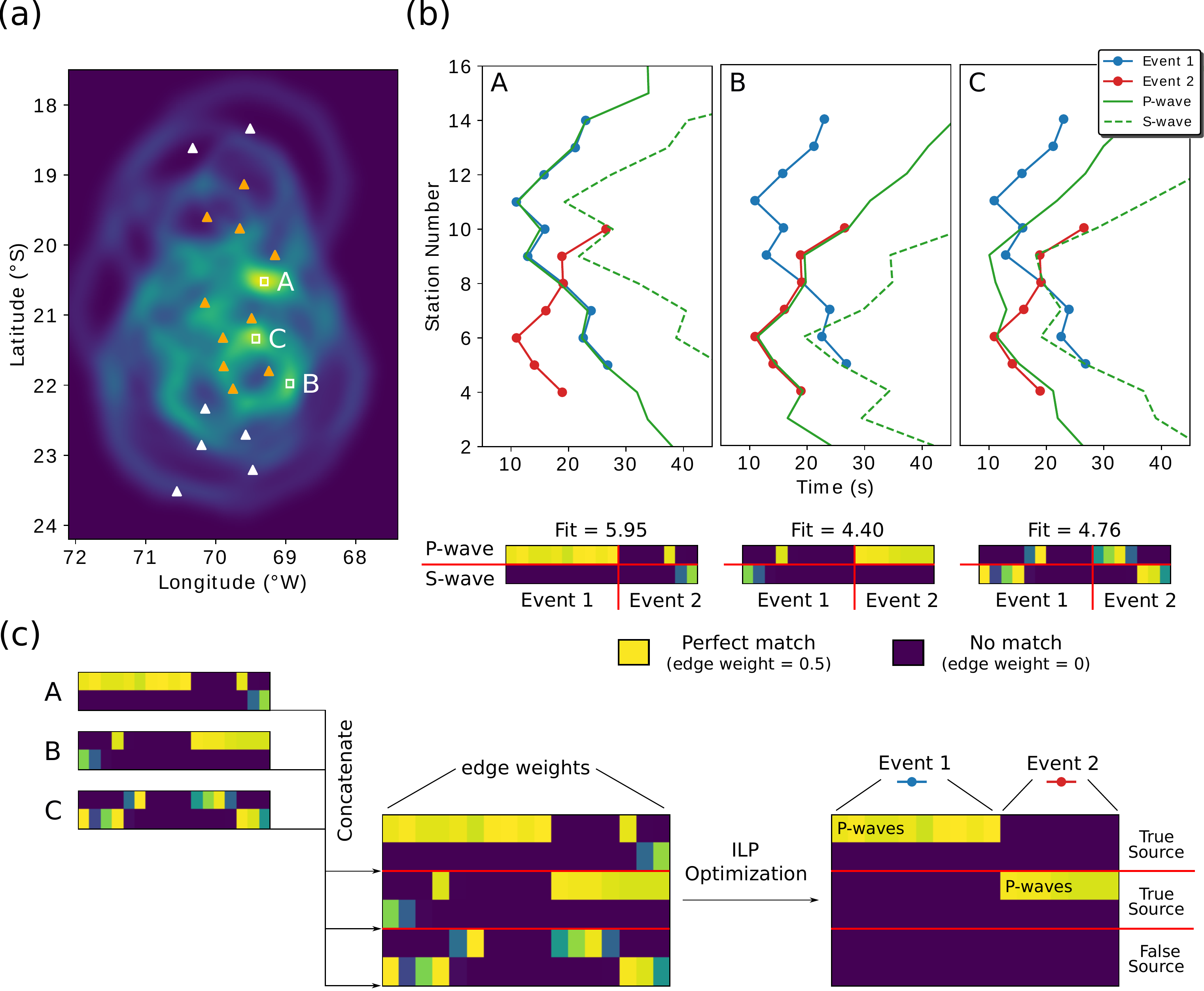}
\caption{Graph construction and ILP optimization result for the two earthquake synthetic example of Fig. \ref{fig:Backprojection}. (a) The BP-space of the combined dataset, as shown in Fig. \ref{fig:Backprojection}d. True sources are marked A, B, and a false candidate source is marked C by white boxes. (b) The predicted arrival times of P- and S-waves for each of the three candidate sources (A, B and C), overlaying the observed arrival time dataset. Below the arrival time data, the measured edge weight matrices are shown for each source-arrival pair, and each phase type. In these matrices, arrivals are arranged left to right by the first earthquakes arrivals, followed by the second earthquakes arrivals. Above each matrix the sum of all edge weights contained in the matrix is given. (c) A schematic of the ILP optimization routine applied to the weight matrices shown in (b). By solving the constrained optimization problem (equation 6), the resulting solution identifies the two correct sources, and the correct phase assignments of each arrival.}
\label{fig:GraphOptimization}
\end{figure}

Given the properties of the source-arrival graph outlined, it is clear such a graph will contain useful information for interpreting the results of backprojection (Fig. \ref{fig:GraphOptimization}). For instance, for an arrival dataset of a single earthquake and its associated collection of candidate sources obtained from discrete backprojection, $\{\tilde{\mathcal{D}}_{(\vect{X},t)}, \mathcal{S}\}$, the true source in $\mathcal{S}$ will have large edge weights to all of the arrivals in $\tilde{\mathcal{D}}_{(\vect{X},t)}$, while each false candidate source will have large edge weights to only a subset of arrivals (each candidate source being connected to different subsets of the full set of all arrivals). Moreover, since each edge weight is recorded as a phase-dependent measurement, the true source will have large edge weights of phase type $r = 1$ to the P-waves in the dataset, and large edge weights of phase type $r = 2$ to the S-waves in the dataset, but not vice-versa. In the following section we describe a constrained optimization routine that can be applied to arbitrary $w_{ijkl}^r$ graphs, which can extract these inferences directly, and which also generalizes to multiple earthquakes being contained in the dataset, and other complicating factors such as missing arrivals and false arrivals.

\subsection*{Competitive Assignment}
\label{subsec:CA}

When viewed on short time scales, the source-arrival graph $w_{ijkl}^r$ generally has a highly entangled set of edge connections, since any number of arrivals may coalesce to cause a source to trigger (i.e., to exceed the threshold for declaring a candidate source), and each arrival may itself contribute to several candidate sources. Our optimization scheme imposes constraints on the connections in $w_{ijkl}^r$ that force arrivals to be linked to at most one candidate source (using only one phase type), while maximizing the number of source-arrival edges retained and applying a negative penalty for all retained sources from $\mathcal{S}$. This causes candidate sources to compete with others to obtain the most connections as possible, and results in many candidate sources being discarded, so that only an optimal (small) suite of inferred true sources are retained. In the trivial case of a single earthquakes dataset and its collection of candidate sources, such an optimization would result in the one true source `winning' the connections to each arrival, and all of the false candidate sources being discarded, since this would maximize the number of assigned arrivals while minimizing the number of active sources. The success obtained on the trivial problem, however, extends to more complicated cases by use of the full optimization algorithm, referred to as `competitive assignment', that we now define.

The optimization problem is defined as a constrained integer linear programming problem (ILP) applied directly to the graph edge weights contained in $w_{ijkl}^r$. ILP's are a type of linear optimization problem in which the solution vector, $x$, is composed of all integers \cite{nemhauser1988integer, schrijver1998theory}, and in this case is binary and has an entry for each edge in the graph specifying whether to keep ($x_i$ = 1) or delete ($x_i$ = 0) that edge. The optimization seeks to maximize $c^Tx$, in which vector $c$ is comprised of all of the edge weights in $w_{ijkl}^r$, with a constraint matrix $A$ acting on $x$ that ensures that:

\begin{itemize}
    \item[(1)] each arrival is assigned to (associated with) at most one source, and one phase type,
    \item[(2)] each station contributes at most one of each phase type to each source, and
    \item[(3)] a user-specified penalty, $-\phi$, is applied for each source that retains $> 0 $ connections.
\end{itemize}

\noindent The competitive assignment problem can then be written succinctly as:

\begin{subequations} \label{eq:Competitive Assignment}
\begin{align}
    \max\limits_x \; &c^Tx \\
    \text{s.t.} \; &Ax \leq b \\ 
         &x \in \{0,1\} \\
    \{A,b,c\} & \xleftarrow[]{Algs. \; S1 - S3} \{\mathcal{D}, \mathcal{S}\} \; ,
\end{align}
\end{subequations}

\noindent where the precise development of the constraints, $\{A,b\}$, optimization vector, $c$, and how to read the result from the binary solution vector $x$ only involve an indexing scheme. The scheme can be automated for any given dataset, $\mathcal{D}$, and set of candidate sources, $\mathcal{S}$. The notation in equation (6d) indicates that Algorithms S1-S3 summarize the steps of this scheme, and details may be found in the Supplemental Material. The binary solution vector $x$ obtained from maximizing (\ref{eq:Competitive Assignment}) encodes the assignment of all arrivals to at most one source, the phase type of their assignment, and the `active' and `inactive' state of all candidate sources. Arrivals not assigned to a source are flagged as possible false arrivals. The solution represents a structured and physically realistic subgraph that's extracted from the input graph that maximizes the number and strength of all source-arrival connections retained, while minimizing the number of active sources needed to do so.

The competitive assignment algorithm can succeed for even complex arrival and candidate source datasets (and their associated graphs) for intuitive reasons. If multiple earthquakes are present in $w_{ijkl}^r$, optimizing (\ref{eq:Competitive Assignment}) will encourage the identification of all of the true sources, since doing so will enable the assignment of the most arrivals. Since each arrival can only be assigned to one source, false sources are unlikely to be retained, since all of the arrivals that originally caused constructive interference at the false sources in $C(\vect{X},t)$ will instead be assigned to their true source, and the cost penalty $-\phi$ will outweigh the benefit of keeping the false sources (even if, for instance, a few arrivals happen to be left over and could otherwise still be assigned to those false sources). Because the edge weights are measured as phase-dependent quantities, assigning each arrival to their true source with the correct phase assignment will maximize (\ref{eq:Competitive Assignment}) more than if an incorrect phase assignment is used. In addition, underneath the backprojection mapping, most false arrivals will not align with the moveout of a true source (as long as $\sigma$ is not too large), and hence (true) source-(false) arrival edge weights will be predominantly zero, and the majority of false arrivals will not be assigned to any source\footnote{Most ILP solvers will allow a zero-weight assignment (as this neither hurts or improves the optimization), and hence, all zero edge weights should be adjusted to a small negative value before optimizing (\ref{eq:Competitive Assignment}) so that false arrivals will be explicitly ignored.}. 

In Fig. 3, the graph construction and ILP optimization steps are shown for three representative candidate sources (local maxima) detected in the BP-space, for the same example of two earthquakes shown in Fig. 2. In this example, the ILP optimization resolves the ambiguity of the BP-space (Fig. 3a) to determine the correct number of sources, and the phase assignments of each arrival. While the arrivals are all treated as having an unknown phase type during the backprojection step, the solution determines that all observed arrivals are P-waves. Notably, the false candidate source (marked C in Fig. 3a) has higher BP coherency than a true source (marked B in Fig. 3a), yet the optimal solution of the ILP correctly identifies source C as being false, and sources A and B as being true. This is possible because the simultaneous joint inversion for all source-arrival assignments reveals that the maximum number of source-arrival assignments is achieved by retaining sources A and B, rather than retaining source C (in combination with either source A or B), and letting source C split the arrival assignments between the two earthquakes.

\subsection*{Continuous-Time Applications and \\ Computational Efficiency}

The number of candidate sources triggered in discrete backprojection has a strong effect on the computational efficiency, as well as the memory requirements of the method. As the duration of the dataset $\mathcal{D}$ increases and more earthquakes are included, inevitably the number of triggered sources increases without bound and practical questions surrounding how to apply the method to continuous-time applications (without artificially subdividing $\mathcal{D}$) arise. A related consideration is whether any of the hyperparameters used in the method should change as seismicity conditions vary. To address both issues and maintain as automated a method as possible, we introduce a few modifications and intermediate steps (all of which are outlined in Algorithm 1 for reference). Specifically, we adaptively choose the triggering ($\gamma$) and cost ($\phi$) hyperparameters as the rate of observed seismicity (over a time-window) changes, and we also introduce a few graph-based tools to reduce the complexity of the graphs passed to the competitive assignment optimization routine. The former ensures that increased rates of seismicity don't blow up the number of false candidate sources triggered, while the later steps are beneficial for keeping the graphs competitive assignment is applied to small and tractable. Most ILP solvers have computation time that scales with $\bigO(n^2)$ for $n$ nodes, and it is thus computationally advantageous to apply competitive assignment to as sparse and small of graphs as possible.

\subsubsection*{Adaptive Triggering Threshold $\gamma$ and Cost Term $\phi$}

During standard operational settings choosing a fixed triggering threshold $\gamma$ for all time-intervals is usually appropriate. However when seismicity rates change during earthquake swarms or aftershock and foreshock sequences (among other possible situations), the number of triggered sources may increase with $\bigO(e^n)$ for $n$ earthquakes and soon become computationally prohibitive. The cause of this is that the base-level of the $C(\vect{X},t)$ space increases as more spurious arrivals are included in $\mathcal{D}$. To maintain a computationally efficient algorithm, it's helpful to increase $\gamma$ proportionally to the background coherency level in $C(\vect{X},t)$, such that the triggered candidate sources still distinctly stand out above the background level of coherency.

To implement a simple adaptive choice for $\gamma$ we choose $\gamma$ to be linearly proportional to the average rate of arrivals occurring across the network in a time-interval. That is, we set

\begin{equation}
    \gamma = \frac{1}{N}(m_{\gamma}r_{obs} + b_{\gamma}) \; ,
\end{equation}

\noindent where $r_{obs}$ is the average rate of arrivals per station, per day, and the coefficients $m_{\gamma}$ and $b_{\gamma}$ are chosen to maintain a balance between accuracy and efficiency. In our application we have found that even a small linear scale of $(m_{\gamma}, b_{\gamma}) =$ (0.00122, 4.9), allowing $\gamma$ to adapt between 6$/N$ and 7$/N$ as $r_{obs}$ changes between 900 and 1800 is sufficient to allow the method to run for either modest or high rates of seismicity with little changes to the resulting computation time or memory cost of the method. After $\gamma$ is decided, we set $\phi = N\gamma - 0.5$ in all cases. The window size over which to compute $r_{obs}$ is subjective, however in our application we compute $\gamma$ over daily seismicity rates.

\subsubsection*{Disconnected Subgraphs}

Mathematical graphs can be separated into distinct subgraphs by searching for populations of nodes that are entirely disconnected from other populations of nodes. When $\mathcal{D}$ includes arrivals spanning a time-interval many times longer than travel times allow, typically $w_{ijkl}^r$ will be highly sparse and contain subsets of nodes that are disconnected from one another (so long as $\gamma$ is not too small). Since these divisions already separate groups of sources/arrivals from one another naturally, we extract all such disconnected subgraphs from $w_{ijkl}^r$ and run the competitive assignment and graph pruning subroutine (described next) on these subgraphs individually.

\subsubsection*{Graph Pruning with the Markov Clustering Algorithm}

To improve the efficiency of competitive assignment (which may become slow when an input graph has, e.g., $>$ 50 sources $\times$ 150 arrivals), we use a graph theory based tool, known as the Markov Clustering Algorithm (MCA), to discard a number of sub-optimal sources contained in $\mathcal{S}$ prior to running competitive assignment. This reduction is achieved by identifying which sources in $\mathcal{S}$ are most likely subsidiary local maxima to different, more optimal sources in $\mathcal{S}$. This information can itself be revealed by a particular consideration of the observational backprojection space, $C(\vect{X},t)$, and theoretical BP-space, $\tilde{C}_{(\tilde{\vect{X}})}(\vect{X},t)$, for a given dataset $\mathcal{D}$ and set of candidate sources $\mathcal{S}$.

To explain the relationship between MCA, $\tilde{C}$, $C$, $\mathcal{S}$ and $\mathcal{D}$, we briefly describe the essence of the MCA algorithm (and also define it precisely in Algorithm S8). MCA is a clustering algorithm applied to stochastic matrices. Stochastic matrices are matrices which define the discrete transition probabilities between any two nodes in a graph when following a random walk. On a graph, a random walk is the trajectory a particle takes when iterating a random sequence of transitions between its current node location to any of the other nodes it has edge connections with, where the individual discrete transition probabilities to neighboring nodes are equal to the edge weights between those nodes. The purpose of MCA is to identify subsets of nodes (or node populations) in a graph that are more interrelated than others, as encoded by the structure of edge weights describing the transition probabilities between the nodes in the graph. Intuitively, if two populations of nodes have many interconnecting edges and only a few weak connections between the two populations, then it is apparent that random walks spend a significantly higher proportion of time walking among either cluster than compared with transitioning between the different clusters. MCA is simply an efficient analytical method which can be applied to stochastic matrices to extract these type of inferences directly (however, it generalizes to any number of clusters). In addition, it also identifies the centroid of each cluster; i.e., the fixed-point node that the rest of the nodes in the cluster are most likely to visit when following a random walk.

This algorithm can be applied to identify which sources in $\mathcal{S}$ are sub-optimal and enable the simplification of $\mathcal{S}$. The stochastic matrix we use in MCA is a source-source indexed graph and is given by

\begin{equation} \label{eq:Stochastic Matrix}
    H_{ij} = \tilde{C}_{(X_i)}(\vect{X}_j,t_j - t_i) \times C(\vect{X}_i,t_i) \; ,
\end{equation}

\noindent which defines the transition probabilities to account for two principles: (1) the transition probabilities are high between pairs of sources $\{(\vect{X}_i,t_i)$, $(\vect{X}_j,t_j)\} \in \mathcal{S}$ whenever the theoretical pairwise BP-space value of these sources is high (which is a symmetric quantity and observation data independent), and (2) the transition is asymmetrically skewed towards the source of higher observational BP coherency. This type of stochastic matrix encourages transitions between sources that are intrinsically likely to be related (as defined by $\tilde{C}$), while biasing transitions away from sources of low coherency towards sources of higher coherency (given by $C$). These properties ensure that cluster centroids returned from MCA should be of higher coherency than other sources contained in $\mathcal{S}$ that they're intrinsically related too, given the theoretical BP-space. In practice, MCA also uses a hyperparameter, $\epsilon$, to help control how coarse or fine the resulting clusters will be. Since we use MCA only to decrease the size of the graphs (rather than solve the detection problem completely), in general we choose a high value of $\epsilon$ (e.g., $\epsilon = 5)$, which promotes returning many clusters, and we discard all sources that are not returned as a cluster centroid. Given the nature of the graphs we typically encounter, even a single earthquake tends to return several cluster centroids, however for some well isolated earthquakes, at times only a single (correct) cluster centroid (or source) for a single earthquake is returned. In either case, we run competitive assignment on the returned reduced graphs, since competitive assignment can enforce physical constraints, while MCA cannot.

\subsection*{Waveform Arrival Picking: Norm Picker}

We use a single station energy-based picking method to pick arrivals from raw seismograms in our application to data from the IPOC network. We call the algorithm the `Norm Picker' and it is designed to reveal onsets of energy much like the STA/LTA \cite{allen1978automatic}. The method is a recursive technique that steps through waveform data over time by a small time step, and at each time step it: (1) slices a finite window ($w_1$) of data from the mean of the envelope of all three-components, (2) normalizes the slice by the RMS, (3) smooths the normalized data by a small moving average time-window ($w_2$), and (4) stacks the slice back onto its corresponding time-interval from which it was sliced. Finally, a (5) moving window ($w_3$) fit of the derivative of this time-series is taken, in which local maxima above a threshold are marked as arrival picks. The algorithmic steps are defined in Algorithm S9, and examples of waveform picks are given in Fig. S2. We used the Norm Picker algorithm in our application to northern Chilean seismicity primarily because of its ease of use and its ability to pick relatively low-SNR P- and S-wave arrivals, however other recent developments in arrival picking (e.g., \citealt{perol2018convolutional, ross2018generalized, wu2018deepdetect, zhu2018phasenet}) could also be used with our association technique.

The reason this picking algorithm is effective at detecting low-SNR arrivals, as well is being relatively insensitive to coda-waves masking S-wave arrivals, is that it adapts a local view of the seismicity depending on how much energy is arriving at any given time. By normalizing the envelope of a waveform by its RMS, where $w_1$ is usually taken a few times longer than the duration of a typical earthquake, a high-SNR arrival will be normalized to a similar dynamic range as a low-SNR arrival will be. When a low-SNR arrival is contained inside the window $w_1$, it will be amplified by the RMS normalization, helping to enable its detection. The last step, which takes a moving window fit to the derivative of stacked RMS-normalized envelope windows, ensures that only when energy is sustained over a time-interval on the order of $w_3$ will it have a high derivative value, which eliminates detecting spurious spikes of energy that do not have a substantial duration in time. Furthermore, the derivative detects the onset of energy, which helps localize the arrival time.

\subsection*{Earthquake Characterization}

Once arrival associations and phase types have been determined, earthquakes can be relocated with standard techniques \cite{tarantola1982generalized, lomax2000probabilistic}. In the observational settings that pertain to the IPOC network, we detect earthquakes that are nearby, as well as distant to the seismic network (e.g., $>$ 250 km distance), including many earthquakes at depths $>$ 200 km. As a result, we have found that the IPOC network is poorly suited to accurately constrain earthquake locations, particularly along the plate interface to the east of the network (even when both P- and S-wave detections have been made), largely because the earthquakes are far away from the seismic stations and the velocity model is poorly known in this region. For the purposes of developing a catalog that remains comparable to existing works \cite{bloch2014high, kato2014multiple, leon2016diversity, sippl2018seismicity}, we chose to relocate events in the NonLin Loc Bayesian framework \cite{lomax2000probabilistic}, in which we implement a prior on the source locations to bias locations closer to the plate interface, which is consistent with the predominant distribution of earthquake locations reported previously \cite{sippl2018seismicity}. Specifically, we define the (unnormalized) prior likelihood of source locations as

\begin{equation} \label{eq:Prior}
    \theta(\vect{X}) = \exp\Bigl(\frac{-||\vect{X} - \vect{X}_{plate}||^2_2}{2\sigma_{\vect{X}}^2}\Bigr)
\end{equation}

\noindent where $\sigma_{\vect{X}}$ is set as 15 km, and $\vect{X}_{plate}$ represents the nearest point on the Slab 1.0 Earth model of the Nazca Plate \cite{hayes2012slab1} to the coordinate $\vect{X}$. The ${||\cdot||}_2$ operator in (\ref{eq:Prior}) represents euclidean distance between coordinates in a Cartesian (rather than elliptical) coordinate system. 

Following estimating earthquake locations with maximum likelihood estimation (equation S4), earthquake origin times are estimated with analytical formula (equation S5), and magnitudes are estimated with a local magnitude scale (equation S6) calibrated to produce magnitudes comparable with the Chilean Centro Sismol\'ogico Nacional (CSN) catalog. As shown in the Application section, the use of a prior succeeds in resolving the intrinsic uncertainties of locations (primarily of depths) to the east of the seismic network, while also not having an overwhelming influence on resulting locations in other regions, since we still detect many shallow earthquakes located near the surface, away from the plate interface, in locations previously reported to have seismicity.

\section*{SYNTHETIC TESTS}

We conduct a set of synthetic tests to evaluate the performance of our method under a range of complexity cases. Specifically, we test the performance of the algorithm as the rate of earthquakes ($r_e$), the rate of false arrivals ($r_f$), and the uncertainty on arrival time picks ($\sigma_a$) all change. To generate synthetic arrival time datasets $\mathcal{D}$ that resemble true observational settings, we use the following procedure: 

\begin{itemize}
    \item[(1)] Choose $r_e, r_f,$ and $\sigma_a$
    \item[(2)] For a time window $T$, sample the number of earthquakes from a Possion distribution with rate $r_e \times T$
    \item[(3)] Sample earthquake hypocenters from $\theta(\vect{X})$
    \item[(4)] Sample earthquake origin times from a Uniform distribution over interval $[0,T]$
    \item[(5)] Calculate theoretical times of all arrivals, and add additional error sampled from a Laplace distribution with mean 0 and scale $\sigma_a$
    \item[(6)] Sample earthquake magnitudes from a Gutenberg-Richter distribution with $b = 1.0$
    \item[(7)] For each earthquake, station, and phase-arrival pair, compute the predicted empirical amplitude of each arrival (using equation S6), and if below a threshold, $A_{min}$, delete the arrival from that station
    \item[(8)] For each station, sample the number of false arrivals from a Poisson distribution with rate $r_f \times T$, and sample random false arrival times from a Uniform distribution over interval $[0,T]$ for each.
\end{itemize}

\begin{figure}[]
\centering
\includegraphics[width=0.90\linewidth]{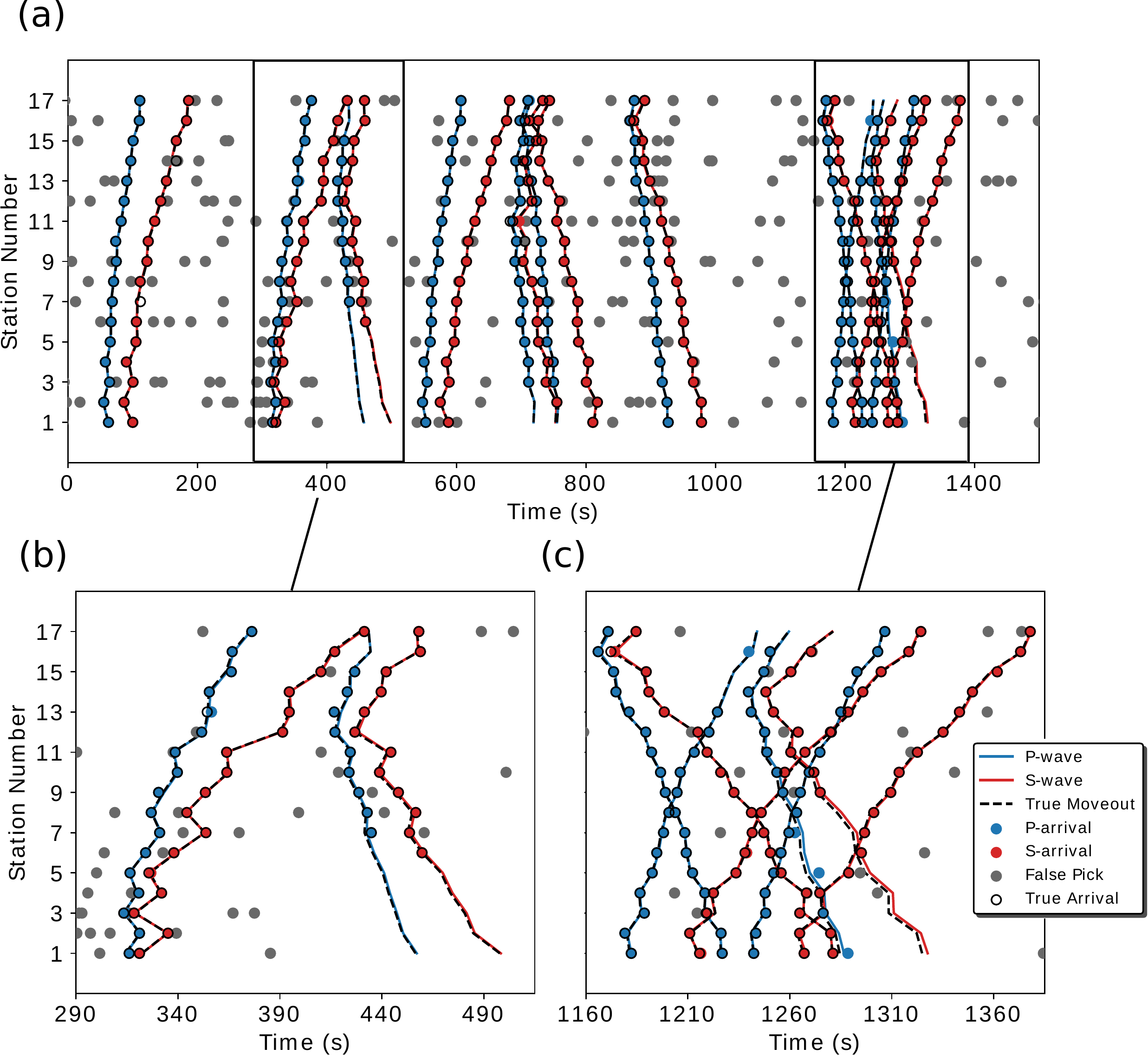}
\caption{Example synthetic test results for earthquake occurring at an average rate of 300 events/day, 300 false arrivals/station/day, and random error on arrival times sampled from a Laplace distribution with mean 0 and scale 1 second. (a) Arrivals from a set of eleven synthetic earthquakes and the predicted events that are matched to them. (b,c) Insets of several earthquakes and source predictions in (a). In (a,b,c) hollow black circles represent missed arrivals, and blue or red circles without black edges represent false arrivals incorrectly associated to an earthquake. Blue and red circles with black edges are correctly associated arrivals. Input data to the algorithm are all arrivals (grey, blue or red) shown in (a), all of which initially have an unknown phase type.}
\label{fig:Synthetic}
\end{figure}

\noindent These steps create realistic arrival time datasets that easily adapt between complex and simple cases by changing $r_e, r_f,$ and $\sigma_a$. In step (7) we use the predicted attenuation of sources to stations to control if arrivals are detected on a station to simulate the real situation in which small earthquakes only produce arrivals on a subset of the nearest stations. The threshold for discarding arrivals, $A_{min}$, is chosen such that $\sim$30$\%$ of all arrivals are missed. For our purposes, we use the same prior, $\theta(\vect{X})$, applied to locate earthquakes to generate the locations of the synthetic sources, which promotes sources that locate nearby the plate interface. It is worth noting that the earthquake origin times and locations are chosen independently, and there is no enforced minimum separation of sources in time or space, which enables many sources to produce overlapping moveouts across the network (Fig. \ref{fig:Synthetic}).

To run the synthetic tests, we set the hyperparameters to the same values used in our application to the real data. We set the RBF-kernel width to $\sigma = 2.8$ s, and let the triggering threshold $\gamma$ follow the linear scaling curve of $\gamma = \frac{1}{N}(m_\gamma r_{obs} + b_\gamma)$, where $(m_\gamma, b_\gamma) = (0.00122,4.9)$ to adaptively adjust for the rate of observed arrivals, $r_{obs}$, and keep the memory requirements of the method low. The cost term in competitive assignment is always set to $\phi = N\gamma - 0.5$. For the experiments, we set $T = 86400$ seconds (one day), and vary $r_e, r_f \in [100, 300, 500, 700]$ events/day, and $\sigma_a \in [1.0, 1.5, 2.0, 2.5]$ seconds. 

\subsection*{Evaluation}

The evaluation of performance is split into several components. For earthquake detections, we count the number of true detections (true positives), missed detections (false negatives), and false predicted sources (false positives). From these values we compute the precision ($P$), recall ($R$), and $F1$ quality metrics \cite{powers2011evaluation}, where precision is given by the ratio of true positives to true positives plus false positives, recall is given by the ratio of true positives to true positives plus false negatives, and $F1$ is a summary statistic given by $F1 = 2 \times (P \times R)/(P + R)$. True detections are marked for all synthetic sources for which the nearest predicted moveout curve is within 6.5 RMS residual of the true moveout curve. False negatives are marked when the nearest predicted moveout curve to a synthetic source has a residual greater than 6.5 RMS. False positives are marked for all other sources we predict that are not matched to a true source. Moveout curve comparisons are used instead of thresholds on nearest location and origin times (between true sources and predicted sources) since moveout curves combine information from both locations and origin times into a single vector, and they are more closely related to the data themselves rather than the specifics of network geometry and source distribution. An additional caveat to the definition of false negatives is that we do not count an earthquake as missing if the attenuation laws in step (7) result in an earthquake having less than $2\times \gamma + 1$ arrivals, since in this case the number of arrivals is so few we don't expect the method to detect the event anyway (recall each arrival can only contribute at most 0.5 to the discrete backprojection metric). This usually occurs for at most $\sim$25$\%$ of the sources in any of the synthetic catalogs. It would be simple and perhaps clearer to delete these sources entirely, however keeping them in the dataset makes the detection problem harder when these arrivals coincide with other larger magnitude sources, which is a phenomenon that happens in the real data and is thus worth including in these tests. We also count the number of correctly associated P- and S-waves and the number of false arrivals correctly identified as false, versus those mistakenly assigned to an earthquake.

\subsection*{Results}

The results of the synthetic tests are summarized in Table 2 and reported in full in Tables S2, S3. We find that as $r_e$ and $\sigma_a$ are increased, the $F1$ detection accuracy statistic varies between 0.99 and 0.93, when taking the median result over all $r_f$ values tested. $r_f$ is found to have a relatively small effect on detection results (Table S2), causing at most a 0.03 variation in $F1$ over the range $r_f \in$ [100, 300, 500 ,700] false arrivals/station/day. In contrast, $F1$ can vary by up to 0.05 and 0.04 units as $r_e$, and $\sigma_a$ vary, respectively, over their testing ranges, $r_e \in$ [100, 300, 500, 700] events/day, and $\sigma_a \in$ [1.0, 1.5, 2.0, 2.5] seconds. In most cases precision is systematically higher than recall (on average by 0.03 units), indicating that the most common type of error is to miss an earthquake rather than create a false source. Example synthetic earthquakes, along with the solutions obtained, are shown in Fig. \ref{fig:Synthetic}. In this example, several earthquakes occur very closely in time, with many overlapping moveouts, yet the solution still resolves this ambiguity.

\begin{table}[t] 

\caption{A summary of the synthetic test results as the rate of earthquakes ($r_e$) and arrival time uncertainty ($\sigma_a$) change. Precision, recall, and F1 of earthquake detection accuracy shown on top row; proportion of correctly associated P-waves, S-waves, and correctly identified false arrivals shown on bottom row. All values are the median value over all false arrival rates ($r_f$) tested. Full set of all results given in Tables S2,S3.}

\begin{minipage}{0.9\linewidth} \centering

\vspace{0.5cm}

\begin{tabularx}{\textwidth}{p{1.75cm} c c c c c c c c c c c c c}

 & & \multicolumn{4}{c}{Precision} & \multicolumn{4}{c}{Recall} & \multicolumn{4}{c}{F1} \\

\cline{2-14}


\multirow{4}{*}{\parbox{0.5cm}{Event/Day, \phantom{.......}$r_e$}} & \multicolumn{1}{c|}{100} & 1.0 & 1.0 & 1.0 & \multicolumn{1}{c|}{.99} & .97 & .99 & .98 & \multicolumn{1}{c|}{.96} & .98 & .99 & .99 & .98 \\
 & \multicolumn{1}{c|}{300} & 1.0 & .99 & .98 & \multicolumn{1}{c|}{.99} & .98 & .97 & .95 & \multicolumn{1}{c|}{.96} & .99 & .98 & .97 & .97 \\
 & \multicolumn{1}{c|}{500} & .99 & .99 & .98 & \multicolumn{1}{c|}{.98} & .98 & .96 & .95 & \multicolumn{1}{c|}{.94} & .98 & .98 & .96 & .96 \\
 & \multicolumn{1}{c|}{700} & .98 & .99 & .98 & \multicolumn{1}{c|}{.96} & .96 & .96 & .95 & \multicolumn{1}{c|}{.91} & .97 & .97 & .96 & .94 \\
 \cline{2-14}
 \addlinespace[0.5ex] & & 1.0 & 1.5 & 2.0 & \multicolumn{1}{c}{2.5} & 1.0 & 1.5 & 2.0 & \multicolumn{1}{c}{2.5} & 1.0 & 1.5 & 2.0 & 2.5 \\ \addlinespace[0.5ex] & & \multicolumn{12}{c}{Arrival Time Uncertainty, $\sigma_a$ (s)}

\end{tabularx}

\vspace{0.5cm}

\begin{tabularx}{\textwidth}{p{1.75cm} c c c c c c c c c c c c c}

 & & \multicolumn{4}{c}{P-waves} & \multicolumn{4}{c}{S-waves} & \multicolumn{4}{c}{False Arrivals} \\


\cline{2-14}


\multirow{4}{*}{\parbox{0.5cm}{Event/Day, \phantom{.......}$r_e$}} & \multicolumn{1}{c|}{100} & .99 & .96 & .90 & \multicolumn{1}{c|}{.85} & .97 & .94 & .89 & \multicolumn{1}{c|}{.84} & .99 & .99 & .99 & .99 \\
 & \multicolumn{1}{c|}{300} & .98 & .95 & .90 & \multicolumn{1}{c|}{.85} & .96 & .93 & .87 & \multicolumn{1}{c|}{.82} & .98 & .98 & .97 & .97 \\
 & \multicolumn{1}{c|}{500} & .97 & .94 & .89 & \multicolumn{1}{c|}{.83} & .95 & .91 & .85 & \multicolumn{1}{c|}{.79} & .97 & .97 & .96 & .95 \\
 & \multicolumn{1}{c|}{700} & .97 & .92 & .88 & \multicolumn{1}{c|}{.82} & .94 & .89 & .84 & \multicolumn{1}{c|}{.78} & .96 & .95 & .94 & .94 \\
 \cline{2-14}
 \addlinespace[0.5ex] & & 1.0 & 1.5 & 2.0 & \multicolumn{1}{c}{2.5} & 1.0 & 1.5 & 2.0 & \multicolumn{1}{c}{2.5} & 1.0 & 1.5 & 2.0 & 2.5 \\ \addlinespace[0.5ex] & & \multicolumn{12}{c}{Arrival Time Uncertainty, $\sigma_a$ (s)}

\end{tabularx}

\end{minipage}

\end{table}

The result accuracy of classifying phase types and identifying false arrivals shows higher variability than the earthquake detection statistics over the range of testing parameters. We find that the proportion of correctly associated P-waves, S-waves, and false arrivals vary between 0.99-0.82, 0.97-0.78, and 0.99-0.94, respectively, when taking the median value over all $r_f$ values tested. For classifying P- and S-waves, $r_e$ and $r_f$ cause variation of up to 0.06 and 0.05 units, respectively, while $\sigma_a$ has the largest effect and can induce changes of up to 0.17 units. This finding, that phase association accuracy is most effected by $\sigma_a$, results because if observed arrival times are too far from their theoretical arrival time, the graph edge weights linking them to their correct sources are often zero. In all cases, P-waves are more accurately identified than S-waves (on average by 0.03 units). False arrival identification accuracy changes by up to 0.06, 0.01, and 0.03 units as $r_e, r_f$, and $\sigma_a$ all vary, respectively, over their testing ranges. Notably, the rate of false arrivals, $r_f$, has the least affect on the ability to accurately identify false arrivals, while $r_e$ has the greatest affect. We interpret the large affect $r_e$ has on false arrival identification accuracy to be that increased numbers of sources make the likelihood for spurious false arrival associations more common.

\section*{APPLICATION: \\ NORTHERN CHILE EARTHQUAKE CATALOG}
\label{sec:Examples}

To demonstrate our method we apply it to 8 years of data from the IPOC network between January 1, 2010 and December 31, 2017. We use data from 17 broadband seismic stations of this network (Table S1), and run the Norm Picking algorithm (Algorithm S9) on the raw seismic data to develop a set of arrival picks before using the association algorithm. Before making arrival picks, seismic data is bandpassed between 5 - 22 Hz to help remove teleseismic arrivals and enhance the detection of small micro-seismic earthquakes. As in the synthetic tests, the number of templates in discrete backprojection is set as $P = $ 500, the RBF-kernel width as $\sigma = 2.8$ s, and the triggering threshold as $\gamma = \frac{1}{N}(m_\gamma r_{obs} + b_\gamma)$, where $(m_\gamma, b_\gamma) = (0.00122,4.9)$, and $r_{obs}$ is calculated as the average number of arrivals per station, per day. The cost term is always set as $\phi = N\gamma - 0.5$.

The source region we consider is between $[$26.5$\degree S$,16.5$\degree S]$, $[$72.5$\degree W$,66.5$\degree W]$, and depths [5 km, -250 km]. For a velocity model we use the trench-perpendicular 2D $V_P$ transect of Model I2 given in \citealt{comte1994velocity}, which we extrapolate north and south throughout the study region. We further assume that $V_s$ scales by 1.76 to $V_p$ \cite{comte1994velocity}. Given this velocity model we calculate $T_i^r(\vect{X})$ on a grid of $(125,85,75)$ (latitude, longitude, depth) elliptically distributed points within the source region using the Shortest-Path Ray Tracing method \cite{moser1991shortest,bai20073d}. A two hidden-layer fully connected neural network is fit to model the travel time grid (Fig. S3) to ensure that $T_i^r(\vect{X})$ is differentiable and continuously known between the original grid-points.

\begin{figure}[] 
\centering
\includegraphics[width=0.90\linewidth]{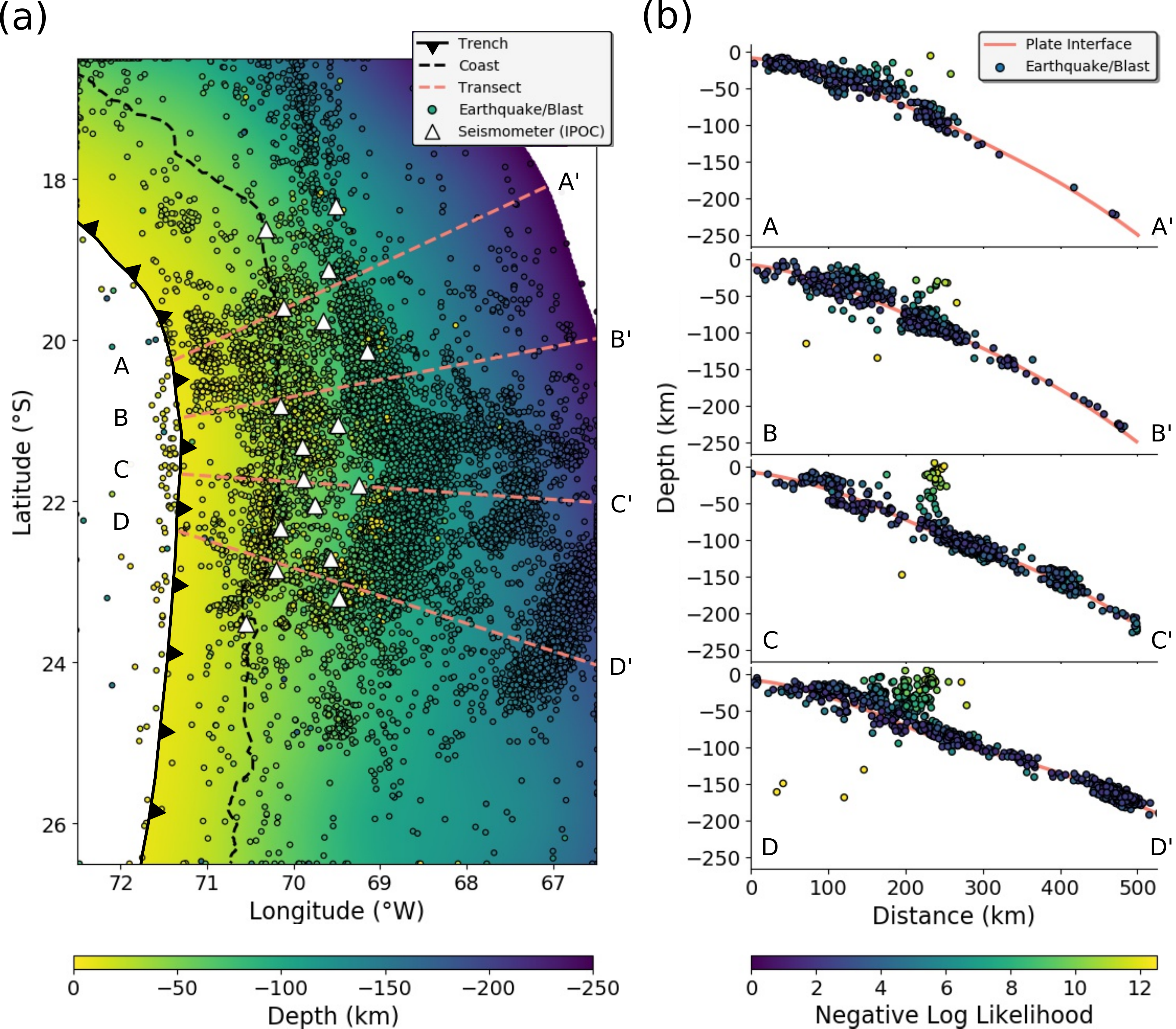}
\caption{Representative set of 16,343 earthquakes from the catalog between the two months of 01/01/2015 - 03/01/2015. (a) Map view of events (circles) colored by depth; background color is the depth to the Slab 1.0 Nazca Plate. Pink lines are cross-section transects labelled A-A', B-B', C-C', and D-D'. (b) Cross-section display of earthquakes within +/- 0.2$\degree$ of transects A-D from panel (a). Circle color represents negative log-likelihood of the source location as calculated by the probabilistic earthquake relocation routine (equation S4). Pink lines denote the depth to Slab 1.0 Nazca Plate along the transects.}
\label{fig:Catalog}
\end{figure}

\subsection*{Results and Validation}

The newly developed catalog we obtain contains a high rate of seismicity, and contains 817,548 detections. We detect on average 279 earthquakes/day (Table 3), and estimate a magnitude of completion of $\sim$M1.8. This is in contrast to previous works, in which 35 earthquake/day were detected in the most complete catalog to date \cite{sippl2018seismicity}, and 7 earthquakes/day are reported in the operational CSN catalog.

\begin{table}[] \centering 

\caption{Catalog comparisons. Time range, number, and event rates listed for CSN, Sippl et al., and this studies catalog. Proportion of matched events between other catalogs and our own with RMS-residual between moveout vectors $<$ 6.5 seconds are listed.}

\begin{minipage}{0.75\linewidth}

\vspace{0.5cm}

\begin{tabularx}{\textwidth}{c c c c c}
\cmidrule(lr){1-5}

 & & CSN & Sippl et al., (2018) & This Study \\

\cmidrule(lr){1-5}
Time Range & & 2007-2017 & 2007-2014 & 2010 - 2017 \\
Number of Events & & 30,796 & 101,602 & 817,548 \\ 
Events/Day & & 7 & 35 & 279 \\
Matched Events & & 98.7$\%$ & 98.3$\%$ & - \\

\cmidrule(lr){1-5}
\end{tabularx}

\end{minipage}
\end{table}

The accuracy of the detections in our catalog are validated in several ways. We (1) confirm the broad distribution of earthquakes is largely consistent with previously reported catalogs, (2) verify that we re-detect a majority of the earthquakes in the CSN and Sippl et al. catalogs, (3) visually assess the accuracy of a set of representative earthquakes, (4) ensure the catalog contains the foreshock and aftershock sequences of the April 1, 2014 Mw 8.2 Iquique earthquake that have previously been reported \cite{kato2014multiple, kato2016accelerated}, and (5) compare the Gutenberg-Richter (GR) curves between this studies catalog and those of Sippl et al. and CSN.

\begin{figure}[] 
\centering
\includegraphics[width=0.90\linewidth]{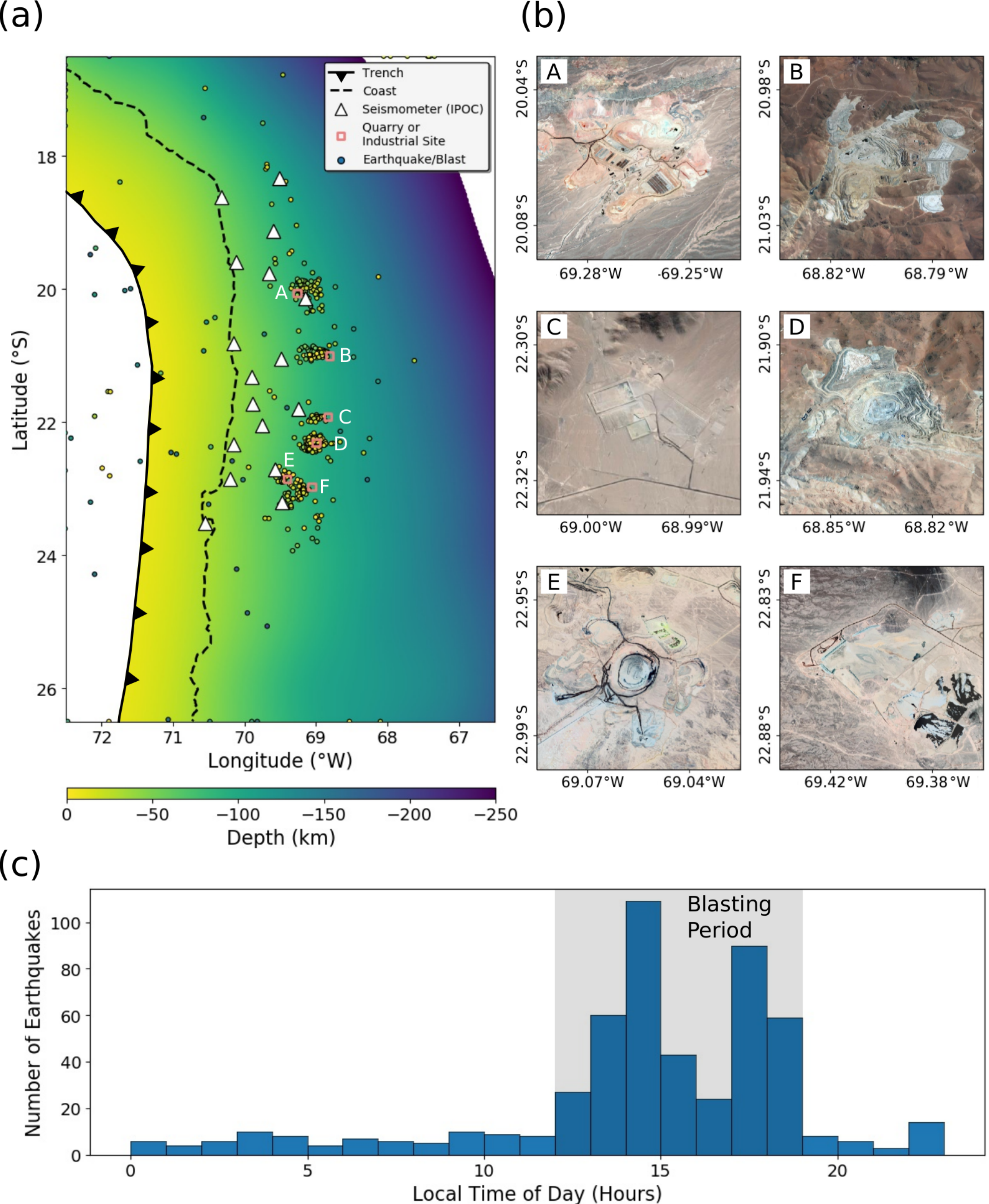}
\caption{Subset of 531 detected earthquakes/blasts from the events shown in Fig. \ref{fig:Catalog}. Events selected have negative log-likelihood location values $>$ 10. (a) Map view of events (circles) colored by depth; background color is the depth to the Slab 1.0 Nazca Plate. Pink squares labelled A-F denote locations of quarry's or industrial sites. (b) Satellite imagery from Google Earth of the six locations marked A-F in panel (a). (c) Histogram of earthquakes/blasts shown in (a) by local time of day (GMT-3). Blasting interval above background level is marked by shaded region.}
\label{fig:Blasts}
\end{figure}

A representative distribution of 16,343 earthquakes from a two month interval (01/01/2015 - 03/01/2015) is shown in Fig. \ref{fig:Catalog}, which shows a spatial distribution broadly in agreement with the CSN and Sippl et al. catalogs (Fig. S4). In particular, the primary bands of trench parallel seismicity, with a transition zone between $\sim$70 - 100 km depth is apparent, and the clustering of deep seismicity ($>$ 125 km) is consistent with that found in both catalogs. Earthquakes at depth largely follow the trend of the subducting Nazca Plate (Fig. \ref{fig:Catalog}b). Some events, however, also locate nearby the surface, with even fewer events locating deep in the mantle to the west. Because of the probabilistic formulation used to locate earthquakes (equation S4) that incorporates the prior, $\theta(\vect{X})$, earthquakes in either of these locations are easily identified by having large negative log-likelihood values (Fig. \ref{fig:Catalog}). The negative log-likelihoods of sources in these regions are intrinsically high (implying a low likelihood value) primarily because the prior likelihood is very small far away from the plate interface.

\begin{figure}[] 
\centering
\includegraphics[width=0.75\linewidth]{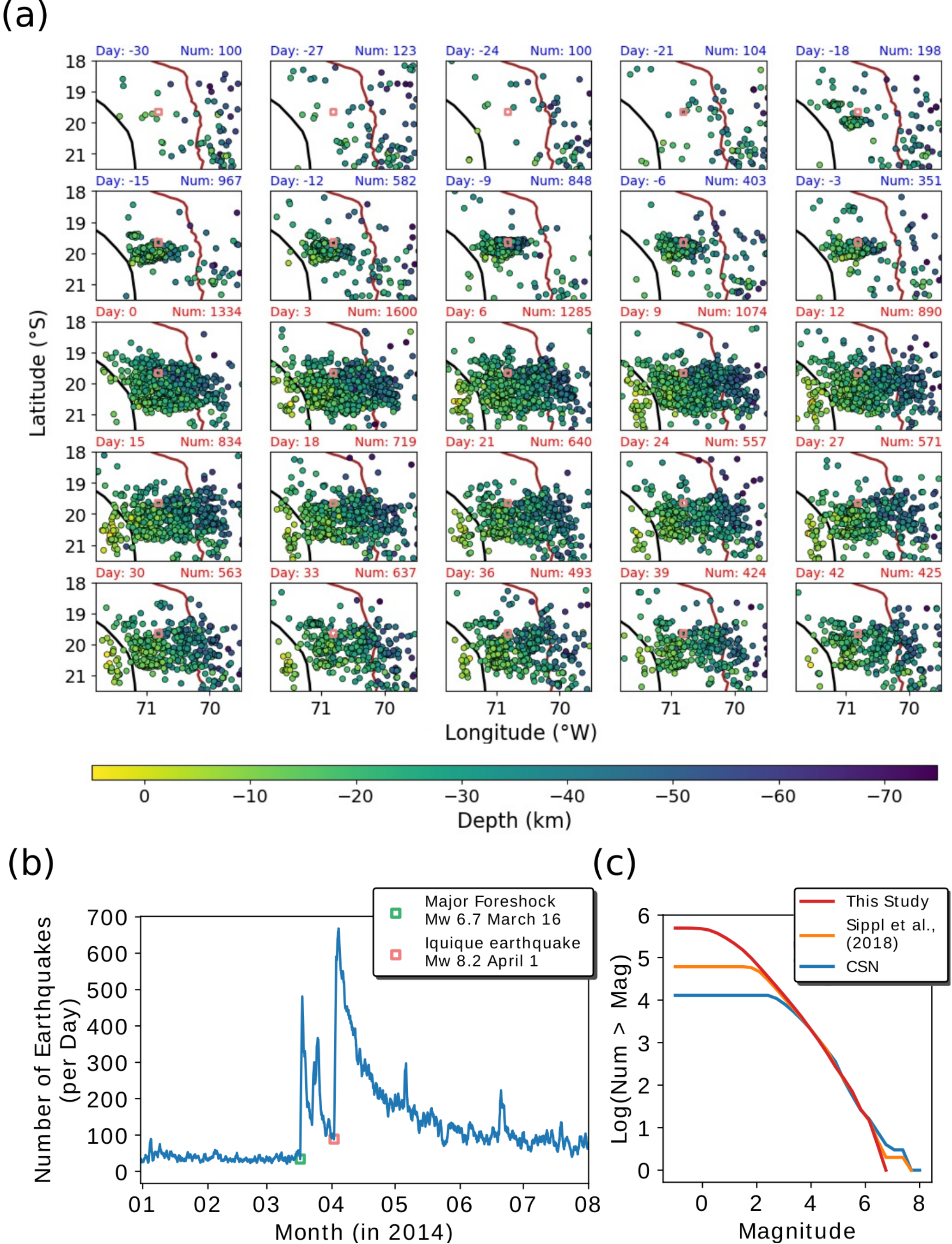}
\caption{Earthquake detections in the temporal and spatial vicinity of the Mw 8.2 Iquique earthquake. (a) Each subpanel plots all epicenters of earthquakes $>$ -75 km depth in the catalog, in sequential non-overlapping windows of three days. Days above each panel list the days before or after April 1, 2014. Numbers list the number of events shown in the subpanel between the day listed and the day of the next subpanel. Pink square marks epicenter of the Iquique earthquake. Other features include: black line (trench), brown line (coast), and blue and red colors of text indicating time intervals before or after the Iquique earthquake, respectively. (b) Earthquake counts between 01/01/2014 - 08/01/2014 in the spatial domain of panel (a) computed over moving 24-hour windows. (c) Gutenberg-Richter magnitude distribution curves for the CSN, Sippl et al., and this studies catalog. GR curves are calculated using overlapping data intervals (between catalogs), as described in the Application section.}
\label{fig:Iquique}
\end{figure}

By examination, we find that sources occurring nearby the surface largely cluster nearby open pit quarries and industrial sites (Figs. \ref{fig:Blasts}a,b). The sources occurring nearby quarries have a non-uniform distribution of origin times with respect to the local time of day (Fig. \ref{fig:Blasts}c), and show clustering between 12 - 9 PM (GMT-3), which is consistent with the predominant blasting times of quarry's as reported in previous studies \cite{sippl2018seismicity}. By visual inspection we have also identified that the anomalous sources occurring deep in the mantle to the west are generally true earthquakes (actually occurring deep in the mantle to the east) that simply have initial locations (following discrete backprojection) to the west because the north-south alignment of the IPOC network results in a strong symmetry, and thus ambiguity, between moveout vectors on either the east or west side of the IPOC network. The initial locations are also far enough from the non-zero regions of $\theta(\vect{X})$ that the gradient based relocation routine cannot relocate the sources (as the gradients are always zero). In a future run of the catalog development we expect that we can account for this discrepancy and relocate these sources by either restricting all templates, $\vect{X}_k$, to non-zero regions of $\theta(\vect{X})$, or by estimating maximum likelihood earthquake locations with a global (rather than local) search procedure. Nonetheless, the fraction of events locating in this region is small enough ($<$ 0.5$\%$) that we do not believe it impacts the majority of our findings.

We compare the three catalogs directly in terms of matched earthquakes and count an earthquake pair as a match (between our catalog and the other catalogs) when the RMS-residual between the nearest matching moveout curve is $<$ 6.5 seconds, in an analogous way as done in the synthetic tests. Doing so shows that we recover $>$ 98$\%$ of the events reported in both the CSN and Sippl et al. catalogs (Table 3). In addition, for all matched earthquakes we calculate the residual differences between source coordinates, origin times, and magnitudes. The residuals are relatively small, with means near zero and standard deviations of $\sim$0.08$\degree$S, 0.18$\degree$W, 21 km, 1.28 seconds, and 0.295 units for latitude, longitude, depth, origin time, and magnitude residuals, respectively (Table S4). Most notably, the largest discrepancy is that for deep earthquakes we locate events $\sim$23 km, and $\sim$13 km shallower than the CSN and Sippl et al. catalogs, respectively, which we infer is caused by using different velocity models between the studies, and potentially by our use of an elliptical (rather than spherical) Earth model.

\begin{figure}[] 
\centering
\includegraphics[width=0.85\linewidth]{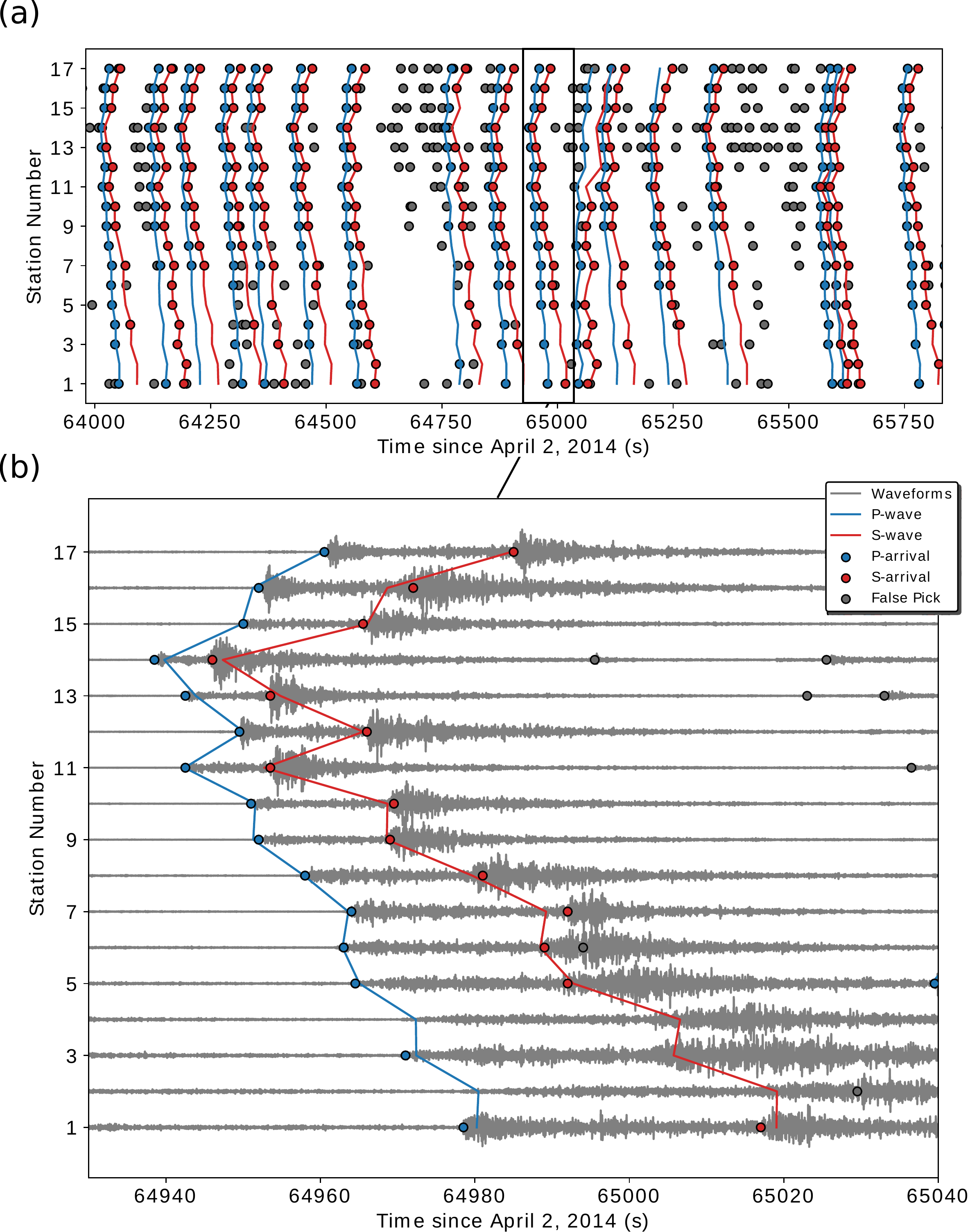}
\caption{Representative set of earthquake detections during April 2, 2014. (a) Arrival times and detections of seventeen earthquakes, including many aftershocks of the April 1, 2014 Mw 8.2 Iquique earthquake. (b) Inset of waveforms from an example earthquake shown in (a). Waveforms shown are HHE channel 50 Hz data bandpassed between 5 - 22 Hz, normalized by their maxiumum amplitudes. In both (a,b), predicted moveouts of P- and S-waves are given by blue and red lines, and predicted P- and S-arrival associations are given by blue and red circles, respectively. Grey circles represent arrivals not associated to a source.}
\label{fig:Real}
\end{figure}

We visually assess the detection quality of a representative set of 300 earthquakes during January 20th, 2014, and assigned each a quality of (1,2, or 3), where to our best judgement, 1 = acceptably accurate, 2 = approximately accurate with some ambiguous arrival assignments or phase classifications, and 3 = likely incorrect set of associations/sources. Carrying out this procedure we found we had a quality distribution of (85$\%$, 11$\%$, 4$\%$) for qualities (1,2,3), respectively, where quality 3 cases nearly always occurred when two or more earthquakes were split incorrectly. By inspection, we have found that the seismicity distribution present during January 20th, 2014 is largely consistent with the seismicity distribution found throughout the catalogs duration, and we do not expect significant deviations from these quality distributions during other time intervals.

We further confirm that the foreshocks and aftershocks of the April 1, 2014 Mw 8.2 Iquique earthquake are detected by our catalog. We detect strong clustering of seismicity near the Mw 8.2 hypocenter for several weeks preceding and following the rupture (Figs. \ref{fig:Iquique}a,b). The number of detections in the restricted interval of [21.5$\degree$S, 18.5$\degree$S], [72$\degree$W, 69.5$\degree$W], and depths [5 km, -75 km] exceeds 600 detections per day during the highest rates of seismicity. These findings give confidence that dynamic sequences of earthquakes (i.e., foreshocks and aftershocks) do not impede the performance of our method, and we are capable of detecting events during both highly activate, and standard background seismicity intervals with no modifications to the method. A representative set of the aftershock detections along with an inset of waveforms is shown in Fig. \ref{fig:Real}.

Lastly, we note that our improved rates of detections compared with previous works results in a GR distribution with lower magnitude-of-completion by $\sim$0.8-1 magnitude unit (Fig. \ref{fig:Iquique}c). Above the magnitudes-of-completion of the different catalogs, the GR curves of Sippl et al., CSN, and our own are largely parallel, indicating a largely consistent set of common detections between the three. To make these GR comparisons fair, we calculated the GR curves using only earthquakes that occurred in spatial and temporal regions that all of the catalogs covered. Specifically, they were computed over the region [25$\degree$S, 18$\degree$S], [71.5$\degree$W, 66.5$\degree$W], depths [5 km, -250 km], and years [2010, 2014].  Further, magnitudes of Sippl et al. and our own are offset by -0.108, and -0.130 units, respectively, to align the GR curves with the CSN catalog between the 3.5 - 5.5 magnitude unit interval.

\subsection*{Computational Cost}

In this application, we found that by using 20 cores of an Intel Xeon 2.40 GHz CPU, which parallelizes over the discrete backprojection, local gradient optimization, and competitive assignment routines (with no GPU acceleration), we process a days worth of seismicity in $\sim$135 seconds, using an average of $\sim$35 GB of RAM during processing. On average, each day 11,000 arrivals across the seismic network are processed, resulting in detecting $\sim$273 earthquakes/day, and discarding 2/3 of all arrivals originally detected by the Norm Picking algorithm as false arrivals (or those caused by magnitude sources small enough that we cannot catalog).

\section*{DISCUSSION}

The problem of earthquake detection is of general interest and a large body of research exists that focuses on the problem of arrival picking \cite{allen1978automatic, gibbons2006detection, yoon2015earthquake, perol2018convolutional, ross2018generalized, zhu2018phasenet}, with comparatively fewer works on arrival associations \cite{bache1993knowledge, johnson1997robust, arora2013net, bergen2018detecting, ross2019phaselink}. With the rapidly improving suite of arrival picking methods, development of effective arrival association algorithms is increasingly important, since associations become more difficult as rates of arrival detections increase with a power law proportional to the arrival picking sensitivity. Issues arise in associating arrivals to their correct sources, as well as in the auxiliary problems of determining the number of sources, determining the phase types of arrivals, and identifying false arrivals. By solving this problem as accurately as possible, high fidelity and low-magnitude-of-completion earthquake catalogs can be obtained. Such catalogs can enhance understanding of many types of seismogenic processes (e.g., spatial-temporal clustering, aftershock and foreshock sequences, dynamic triggering, or long term rate changes). 

In this paper we have presented an algorithm which uses tools from backprojection and graph theory to address the issues inherent to the problem of associations and enable the development of dense catalogs after arrival time picks have been made with a sensitive arrival picking algorithm (e.g., the Norm Picker described herein, or other alternatives \cite{yoon2015earthquake, perol2018convolutional, ross2018generalized, zhu2018phasenet}). We have demonstrated how this technique can have high accuracy on relatively challenging synthetic and real detection problems (Figs. \ref{fig:Synthetic},\ref{fig:Real}), and have also used the method to develop a catalog in northern Chile between 01/01/2010 - 12/31/2017 that has a magnitude-of-completion of $\sim$M1.8, which lowers the magnitude-of-completion by 0.8-1-unit compared with existing catalogs (Table 3, Fig. \ref{fig:Iquique}c). Visual verification of a representative set of newly detected events was used to validate the detections to find that we have an approximately (85$\%$, 11$\%$, 4$\%$) distribution of detection qualities (1,2,3), respectively. Other forms of validation are that we detected the aftershocks and foreshocks of the April 1, 2014 Mw 8.2 Iquique earthquake with no modifications to the method, and we also detected a number of sources nearby local quarrys and industrial sites (Fig. \ref{fig:Blasts}). We re-detected nearly all previously cataloged earthquakes in the most complete catalog of this region to date (Table 3), while increasing the average rate of detection from 35 events/day to 273 events/day. In the application to northern Chile, we found we could process a days worth of seismicity using 20 cores of an Intel Xeon 2.40 GHz CPU in $\sim$135 seconds, enabling the creation of all 8 years of the catalog in $\sim$4.5 days of processing time.

One key component of our method that could be improved would be in generalizing our use of the backprojection metric. Currently, equation (\ref{eq:Backprojection}) assumes that all stations can theoretically detect arrivals of all earthquakes. However, for large aperture networks, the majority of small earthquakes will produce arrivals only on subsets of the networks stations due to attenuation. By backprojecting arrivals from the entire network at all times to all templates, the current metric would allow spurious, unrelated arrivals from distant portions of the network (which line up the moveout of a source but are not physically related) contribute to the detection of a source and be falsely assigned to that earthquake. If it is possible to modify (\ref{eq:Backprojection}) such that all sources only `look' locally for the stations that can potentially observe that source (given its magnitude), then the association method could scale to large aperture networks while maintaining its ability to detect both small and large earthquakes. Additionally, if the directivity of arrivals were known, (\ref{eq:Backprojection}) could be modified to only propagate waves in the direction of the source which would significantly reduce the number of false local maxima in the BP-space. Other places in which the method can be improved are in adaptively adjusting the RBF-kernel widths to account for known variability in travel time uncertainties (between different stations and candidate source regions), and also in incorporating additional information into the graphs prior to applying competitive assignment. For example, edge weights could be weighted by the predicted phase-likelihood of each arrival \cite{zhu2018phasenet}, and if pairs of arrivals are known with high certainty to come from a common source \cite{bergen2018detecting, mcbrearty2019pairwise}, a constraint could be added into the constraint matrix of (\ref{eq:Competitive Assignment}) to force these arrivals to be assigned to a common source. An avenue of development, more generally, may be in constructing arrival-arrival indexed graphs (with edge weights proportional to the likelihood those arrivals are associated) which may contain additional, or complimentary information to be used with the current approach. Such graphs have the advantage that the number of nodes is always bounded by the initial dataset and cannot blow up with increasing numbers of false candidate sources.

The earthquake catalog we have produced in this study may contain valuable data for future studies. A large body of work is now employing machine-learning-based methods that benefit from large training datasets. By containing over 800,000 earthquakes, the catalog presented here may be a valuable source of training data for future machine-learning-based studies. Other possible directions of future research would be to investigate the dynamics of small-scale spatial-temporal interactions of earthquakes, monitor rate changes of seismicity, or investigate for the prevalence of dynamic triggering contained in the catalog.

\section*{CONCLUSION}

Associating earthquake arrivals remains a key component to developing dense earthquake catalogs, and it is crucial that association methods can resolve uncertainties such as unknown number of sources, unknown phase types, false arrivals, uncertainties on arrival time picks, travel time calculations, and the situation where several earthquakes occur nearby in time and produce overlapping moveouts across the network. We have shown that by combining tools from backprojection and graph theory a robust solution to all of these problems can be obtained. Our solution solves for all sources and their phase assignments simultaneously, rather than inferring each source sequentially in a greedy fashion, as is common in other association routines. We have demonstrated our method on synthetic tests as well as the development of a new catalog in northern Chile between January 1, 2010 and December 31, 2017 that has a magnitude-of-completion of $\sim$M1.8. The new catalog in northern Chile contains $\sim$8 times more earthquakes from previously reported catalogs in this region, and may contain valuable insights into seismogenic processes occurring at the active Nazca-South American plate subduction zone yet to be discovered.
\section*{DATA AND RESOURCES}

This work included data from the CX seismic network, obtained
from the GFZ Data Service. The catalog developed in this work is given in the Supplemental Materials. All data processing is done in MATLAB and Python. In the future, we plan to release an open-source Python implementation of the association and arrival picking methods described in this work.

\section*{ACKNOWLEDGMENTS}

This work was funded by Institutional Support (LDRD) at the Los Alamos National Laboratory. We thank Alex Hutko and Ben Baker for helpful feedback on our work. We also thank Bertrand Rouet-Leduc, Claudia Hulbert, Christopher Ren, James Theiler, Nicholas Lubbers and Daniel Trugman for many helpful discussions.

\bibliographystyle{apa}


\newpage

\begin{algorithm} \label{alg:A_1}
\caption{The Association and Detection Method. This flowchart summarizes the essential steps to the method: (1) extracting template moveout vectors, (2) running discrete backprojection and detecting all candidate sources, (3) computing the source-arrival graph edge weights, (4) simplifying the source-arrival graphs by extracting disconnected components, running the graph pruning routine (MCA), running competitive assignment to compute the final set of detections, and (5) estimating earthquake locations, origin times and magnitudes.}
\begin{algorithmic}[1] \vspace{0.2cm}
\Procedure{initialize}{}
\State \textbf{Input: } $\hat{T}_i^r(\vect{X})$ for $\vect{X} \in \Omega_{\vect{X}}$
\State Choose $P$
\State Use K-Means to infer $P$ optimal $\{\vect{X}_k\}$ template coordinates given $\hat{T}_i^r(\vect{X})$
\EndProcedure

\Procedure{run discrete backprojection}{}
\State \textbf{Input: } $\mathcal{D}$ and $\{\hat{T}_i^r(\vect{X}_k)\}$
\State Choose $\sigma$ and $\gamma$ (or adaptively choose $\gamma$ given $\mathcal{D}$)
\State Calculate $C(\vect{X}_k,t)$ using equation (\ref{eq:Discrete Backprojection})
\State Record all triggering times $\{t_{k,l}\}$ where $C(\vect{X}_k,t) \geq \gamma$
\State Convert triggering times into source coordinates; $(\vect{X}_k,t_{k,l} - \min_{i,r} T_i^r(\vect{X}_k)) \in \mathcal{S}$
\State Optionally optimize sources in $\mathcal{S}$ in continuous $\Omega$ with local gradient ascent
\EndProcedure

\Procedure{compute graph}{}
\State \textbf{Input: } $\mathcal{D}$ and $\mathcal{S}$
\State For each source-arrival pair, $(\tau_{i,j},t_{k,l})$, calculate edge weights $w_{ijkl}^r$ using equation (\ref{eq:edge weights})
\EndProcedure

\Procedure{simplify and solve each subgraph}{}
\State \textbf{Input: } $w_{ijkl}^r$
\State Choose $r$ and $\phi$
\State Extract all disconnected subgraphs from $w_{ijkl}^r$
    \ForAll {disconnected subgraphs}
        \State Run MCA with parameter $\epsilon$ to simplify each subgraph
        \State Run CA with paramater $\phi$ on the reduced subgraph and record solution
    \EndFor
\EndProcedure
\Procedure{characterize each earthquake}{}
\State \textbf{Input: } All sources and associated arrivals returned from CA
    \ForAll {earthquakes}
        \State Relocate with probabilistic formulation (equation S4)
        \State Estimate origin time (equation S5)
        \State Estimate magnitude (equation S6)
    \EndFor
\EndProcedure

\end{algorithmic}
\end{algorithm}







\end{document}